\documentclass[aip,jcp,graphicx]{revtex4-1}

\usepackage{graphicx}
\usepackage{dcolumn}
\usepackage{bm}
\usepackage{color}
\usepackage{amsmath}

\begin{document}


\title{Simulation of  Macromolecular Liquids with the Adaptive Resolution Molecular Dynamics Technique} 



\author{J.H. Peters}
\author{R. Klein}
\author{L. Delle Site}
\email[]{luigi.dellesite@fu-berlin.de}
\affiliation{Institute for Mathematics, Freie Universit\"at, Berlin, Germany}


\date{\today}

\begin{abstract}
We extend the application of the adaptive resolution technique (AdResS) to liquid systems composed of alkane chains of different lengths. The aim of the study is to develop and test the modifications of AdResS required in order to handle the change of representation of large molecules. The robustness of the approach is shown by calculating several relevant structural properties and comparing them with the results of full atomistic simulations. The extended scheme represents a robust prototype for the simulation of macromolecular systems of interest in several fields, from material science to biophysics.
\end{abstract}


\maketitle 

\section{Introduction}
\label{sec:introduction}
The Adaptive Resolution Simulation technique (AdResS) \cite{adress-jcp1, annurev} has been developed to efficiently treat only those degrees of freedom which are strictly required for the calculation of certain properties in a molecular simulation. In standard AdResS the simulation box is divided into three regions, an explicit region (or atomistic region), a hybrid or transition region, and a coarse-grained region. In the explicit region molecules have atomistic (or in general high) resolution, in the coarse-grained region they have a low number of degrees of freedom (up until now a spherically symmetric potential acting on the center of mass of a molecule is usually used), and finally in the hybrid region, at the interface between the explicit and the coarse-grained regions, the interactions are controlled by a space-dependent interpolation of the explicit and coarse-grained forces. 
In specific terms the interpolation formula for a molecule $\alpha$ and a molecule $\beta$ is
\begin{equation}
\label{eq:force_interpolation}
F_{\alpha \beta} = w(r(X_{\alpha}))w(r(X_{\beta}))F_{\alpha\beta}^\mathrm{atom} + [1 - w(r(X_{\alpha}))w(r(X_{\beta}))]F_{\alpha\beta}^{cm}
\end{equation}
where $r(X)$ describes the distance of the atom at $X$ from the center of the explicit region (depending on the geometry of the system, this can for example be the spherical distance or the distance along an axis) and 
\begin{equation*}
    w(x) = \begin{cases}
               1               & x < d_{AT} \\
               cos^{2}\left[\frac{\pi}{2(d_{\Delta})}(x-d_{AT})\right]   & d_{AT} < x < d_{AT}+d_{\Delta}\\
               0 & d_{AT} + d_{\Delta}< x
           \end{cases}
\end{equation*}
where $d_{AT}$ and $d_{\Delta}$ are the sizes of the atomistic and hybrid regions, respectively.
The weighting function $w(x)$ smoothly varies from 0 to 1 in the transition region, where a coarse-grained molecule transforms into an atomistic molecule and vice versa.\\ 
 Moreover, to assure thermodynamic equilibrium a thermostat is applied and it has been shown (numerically and conceptually) that very high accuracy of the results can be reached if a one-body force acting on the center of mass of the molecule, called thermodynamic force, is applied \cite{jcpsimon,prlthf}.\\
The thermodynamic force in the form of a gradient of the density has been proven, in a rigorous way, to be a necessary mathematical condition in order to have a grand-canonical-like distribution in the atomistic region \cite{jctchan,prx} and to balance the difference in chemical potential between the atomistic and coarse-grained region \cite{jcpsimon,AgarwalEtAl2014}. At the same time, it was proven that the coarse-grained region can be represented by any generic (non calibrated) molecular model as long as its density and temperature are equivalent to the density and temperature  of the atomistic region \cite{jctchan,prx,AgarwalEtAl2014}. Later on, similar conclusions about the coarse-grained model were reached in the context of an alternative version of AdResS based on the interpolation of potentials instead of forces (H-AdResS); moreover the thermodynamic force is interpreted as the gradient of a one-body potential
which balances thermodynamic differences between the atomistic and coarse-grained region in a global canonical ensemble \cite{PotestioEtAl2013,EspanolEtAl2015}.

\begin{figure}
\includegraphics[width=0.9\textwidth]{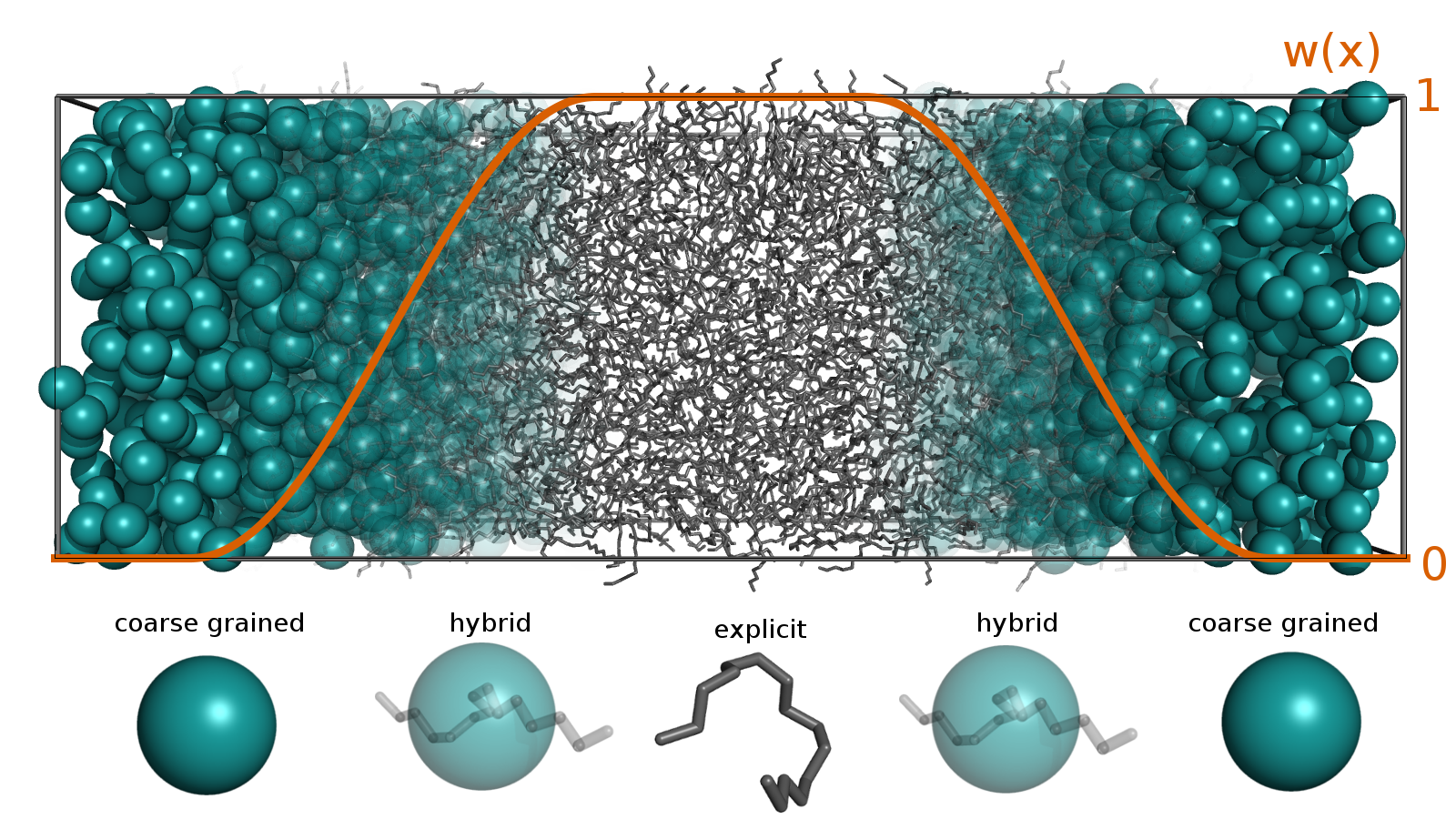}
\caption{\label{fig:system} AdResS simulation box of liquid pentadecane. Overlaid on the structure is the  AdResS weighting function w(x) used in Eq. \ref{eq:force_interpolation}. It is symmetric along the x-coordinate with respect to the center of the box at $x = 10$ nm and defines an atomistic region $w(x) = 1.0$ nm, the two hybrid regions $0.0$ nm $ < w(x) < 1.0$ nm, and the coarse-grained region ($w(x) = 0.0$ nm across the periodic boundary.}
\end{figure}

The method has been successfully applied to several relevant systems, including the solvation of a polymer in a liquid of small molecules \cite{jcppol}, 
pure liquid water \cite{jpcm,jcpwater}, 
mixtures of small molecules \cite{AgarwalEtAl2014} or 
the solvation of large hydrophobic and biological molecules \cite{jcpfuller,matejprot,matejdna,kurtdeb,kurtbio}. 
It also has been extended to systems characterized by quantum delocalization of atoms in space such as liquid parahydrogen at low temperature and liquid water via a path-integral approach\cite{prlado,parah1,parah2,animpi}. 
The conceptual basis of the method has been framed in a grand-canonical-like construction of the hybrid and coarse-grained regions as a large reservoir for the (relatively small) atomistic region which assured conceptual robustness together with technical efficiency \cite{jctchan,prx,njp}.\\
The aim of this work is to go beyond the systems considered so far and further develop the technical capability of the method to enable the change of resolution of large linear chain-like molecules from full atomistic to a much simpler coarse-grained (one spherical or two connected spherical beads) representation. Such kinds of molecules differ substantially from the molecules treated so far in an adaptive fashion, i.e., small molecules with localized intramolecular interactions (usually bond vibrations, if at all) whose shape can be easily described by a coarse-grained spherical model. As a consequence the change of resolution for these molecules (i.e., passing from a sphere to an atomistic resolution) is relatively smooth. Long linear chainlike molecules are characterized by intramolecular interactions which span over a much larger length. Moreover, their shape can be only roughly reduced to a spherical coarse-grained representation (and/or a small number of  connected spheres, as reported later). Above all, chainlike molecules in a liquid are entangled. We consider a molecular liquid to be entangled if the conformational freedom of the molecules is strongly restricted by topological hindrances caused by the surrounding molecules. As a consequence the change of resolution from a spherical coarse-grained representation to a full atomistic representation is not as simple as the case of small molecules. In this paper we propose a technical advancement of the method which allows for an efficient change of molecular representation of large/complex molecules without introducing any physical artifacts in the atomistic region. For simplicity and in order to have a systematic comparison of different chain lengths we have chosen alkane chains, from hexane ($C_{6}H_{14}$) up to pentacontane ($C_{50}H_{102}$), as test systems. For a pictorial representation of (an example of) a system considered see Figure \ref{fig:system}.
In general, the method presented here can be applied to any polymer-like system, from standard polymer melts to systems of biological interest. 
In fact, in current research there are many questions concerning the local microscopic behavior of polymeric/macromolecular systems which could be optimally treated by AdResS (compared to a full atomistic simulation); for example the determination of the local structural arrangement of chains around a small guest molecule \cite{grest}, or in biological-related systems a large number of water molecules, lipid molecules and eventually also proteins, needs to be considered only in a small region where some local details are of interest (e.g., binding processes \cite{ShanEtAl2011,PetersDeGroot2012} or the interaction of antimicrobial peptides and lipid membranes \cite{WangEtAl2014}). In general, local microscopic details can be efficiently resolved by AdResS which would act as a local computational magnifying glass around the region of interest, keeping the rest of the system at lower resolution but at physically consistent macroscopic/thermodynamic conditions.
Finally, it is mandatory to report that recently the case of a melt of star polymers studied with AdResS has been successfully accomplished \cite{stapol}. This is certainly an important supporting argument for the overall robustness of AdResS, and actually strengthens even further the relevance of our results. In fact, star-polymers, despite being large molecules,  can be reasonably represented by a coarse-grained spherical model; moreover their entanglement is only partial (peripheric). This implies that the technique we are proposing here allows us to go beyond the most advanced application of AdResS for macromolecular systems.
The paper is organized as follows: in the next section we provide a description of the technical modifications done to the original AdResS, next comes the sections with results followed by discussion and conclusions.
\section{Coarse-grained Model}
\label{sec:cg}
\subsection{Representing short chains as single spheres}

\begin{figure}
\includegraphics[width=0.9\textwidth]{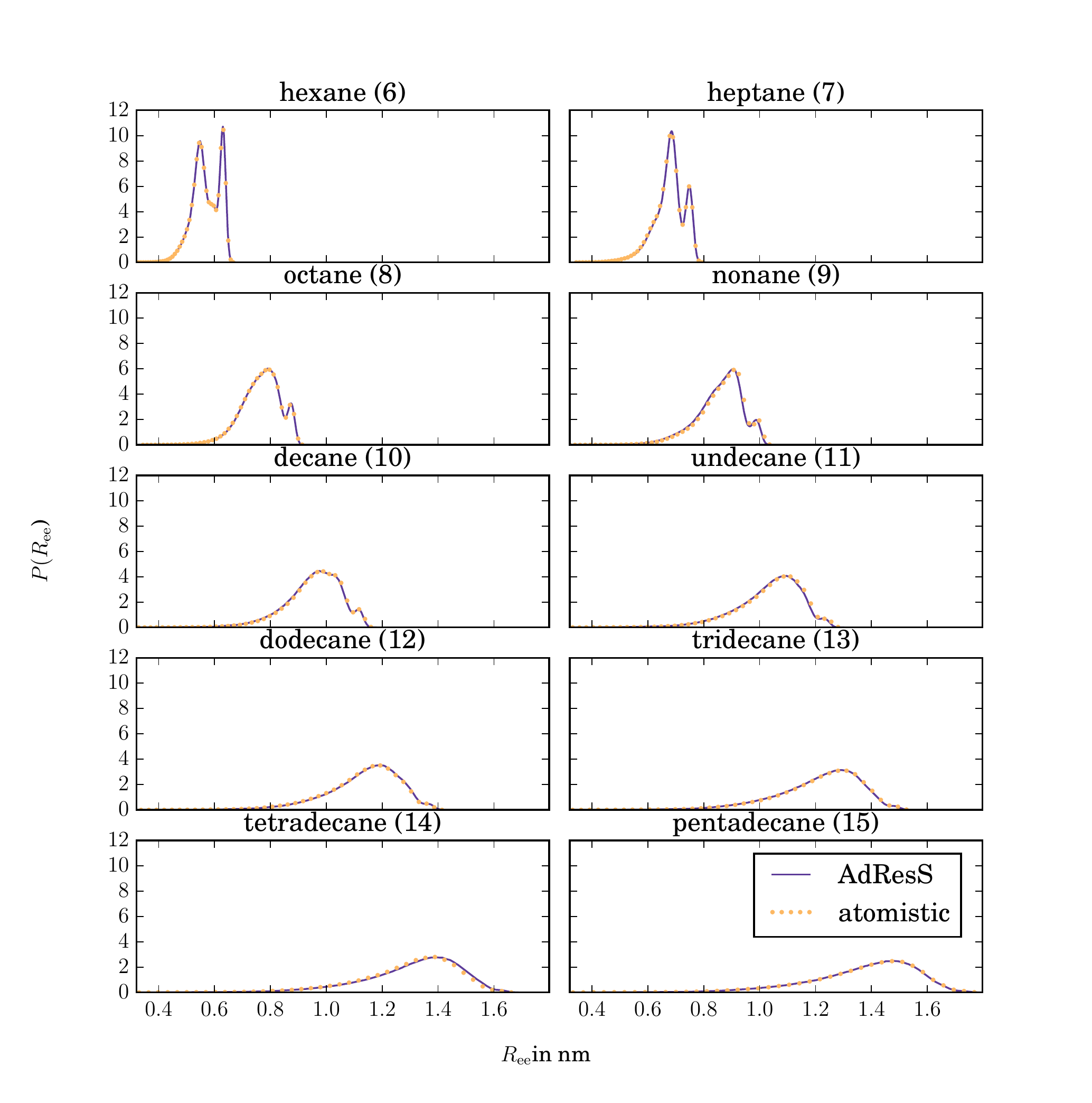}\\
\includegraphics[width=0.9\textwidth]{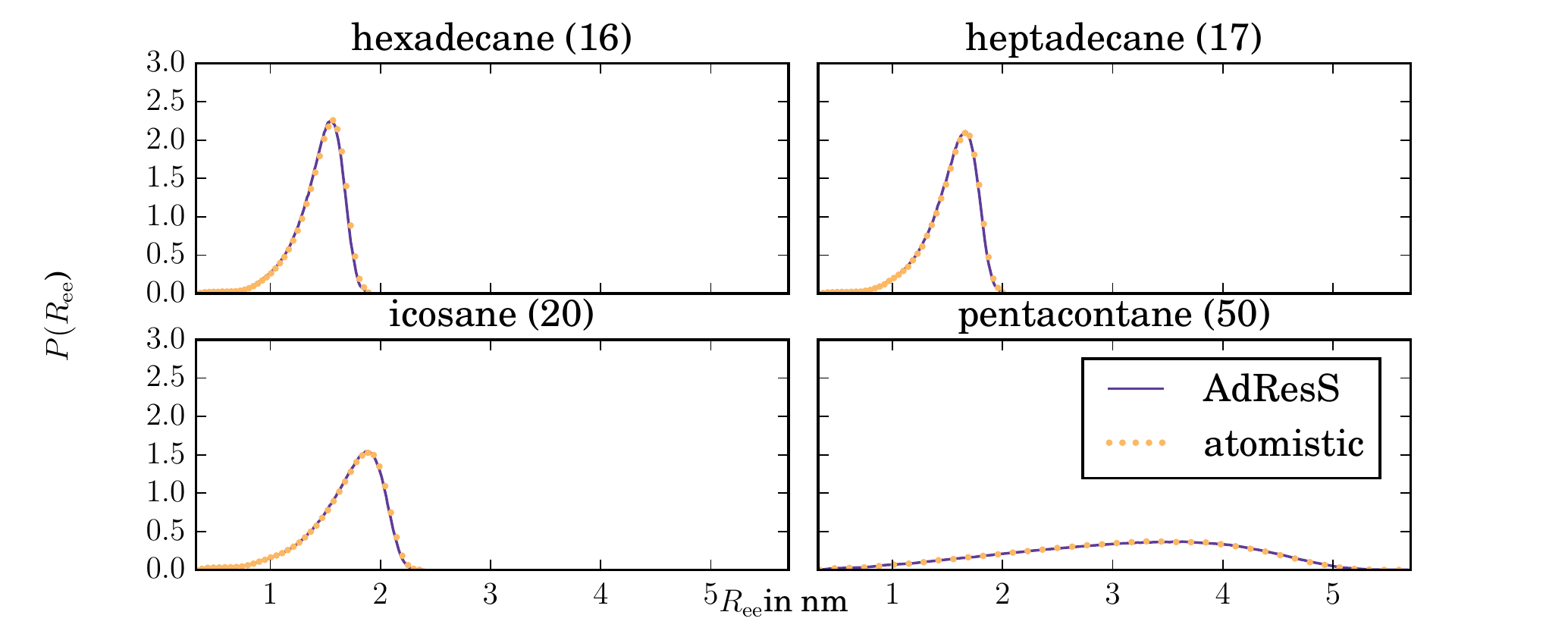}
\caption{\label{fig:PRendend} Distribution of the end-to-end distance, averaged over all molecules in the atomistic region over 9000 frames from a 90 ns simulation after a 10 ns equilibration period. Result from AdResS (blue lines) and full atomistic simulation (yellow dots).}
\end{figure}

The coarse-grained interaction potential was calculated using iterative Boltzmann inversion as provided by the VOTCA toolbox \cite{RuhleEtAl2009}. Coarse-grained beads are defined by the center of mass of the group of atoms they represent. The radial distribution function of these beads was calculated from a 100 ns explicit simulation (detailed simulation parameters in the Appendix). A potential reproducing this radial distribution in the coarse-grained representation was calculated in 300 iterations.
It should be noted that the diameter of the spheres representing the polymer molecules (indicated by the excluded volume in the center of mass radial distribution function depicted in Figure \ref{eq:LennardJones}) is usually smaller than the chain length of the polymer (indicated by the end-to-end distance, Figure \ref{fig:PRendend}). Hence, representing the polymer chain by such a sphere is only a rough but computationally efficient approximation.
As in any (current) AdResS simulation the reduction of degrees of freedom in the coarse-grained region applies only to the intermolecular interactions whose computational cost can be estimated to be between $60$ and $80\%$ of the computer time of a full atomistic simulation of the systems covered here.

The atom-atom interactions between atoms of different chains are reduced to the center-center interaction between the two different coarse-grained beads each representing one polymer, while bonded intramolecular interactions are left untouched. These bonded interactions together with the coarse-grained intramolecular interactions (via GROMACS virtual sites) and the thermostat determine the movement of the molecules (in atomistic representation) in the coarse-grained region.

While this representation originally was adapted to allow an easier implementation of the method, we will now show that this approach offers a promising perspective for the representation of more complex molecules.

\subsection{Using more than one coarse-grained interaction site for longer chains}
\begin{figure}
\includegraphics[width=0.9\textwidth]{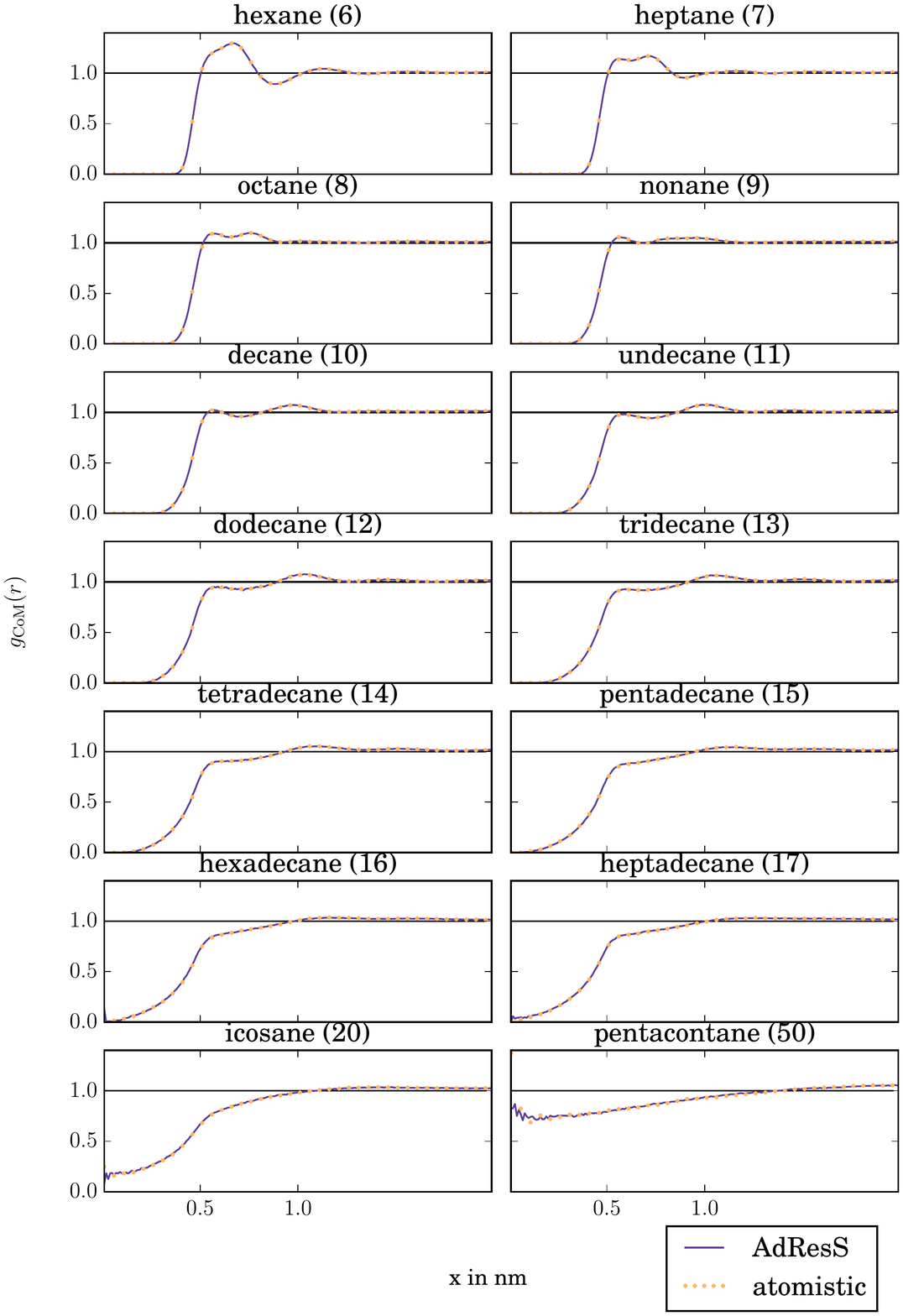}
\caption{\label{fig:rdfCOM} Radial distribution function for the center of mass of the molecules, calculated over 9000 frames from a 90 ns simulation after a 10 ns equilibration period. Result from AdResS (blue lines) and full atomistic simulation (yellow dots). For longer chains, the radial distribution of the molecular centers of mass degenerates, as the center of mass of a long bent polymer chain is not necessarily located near any of the monomers of this chain.}
\end{figure}

For chain lengths of 16 carbon atoms (hexadecane) or longer, the flexibility of the chains results in conformations, where the centers of mass of two entangled polymer chains can coincide (Figure \ref{fig:rdfCOM}) and a coarse-grained representation using only one coarse-grained sphere becomes infeasible. A multibead coarse-grained model has been used in AdResS before to create a polarizable coarse-grained water model \cite{ZavadlavEtAl2015}. In contrast to the supermolecular water clusters modeled in that case, a bond in a coarse-grained model of a linear polymer would have to be mapped onto atomistic bonds during the transition, resulting in possible numerical instabilities.

\begin{figure}
\includegraphics[width=0.3\textwidth]{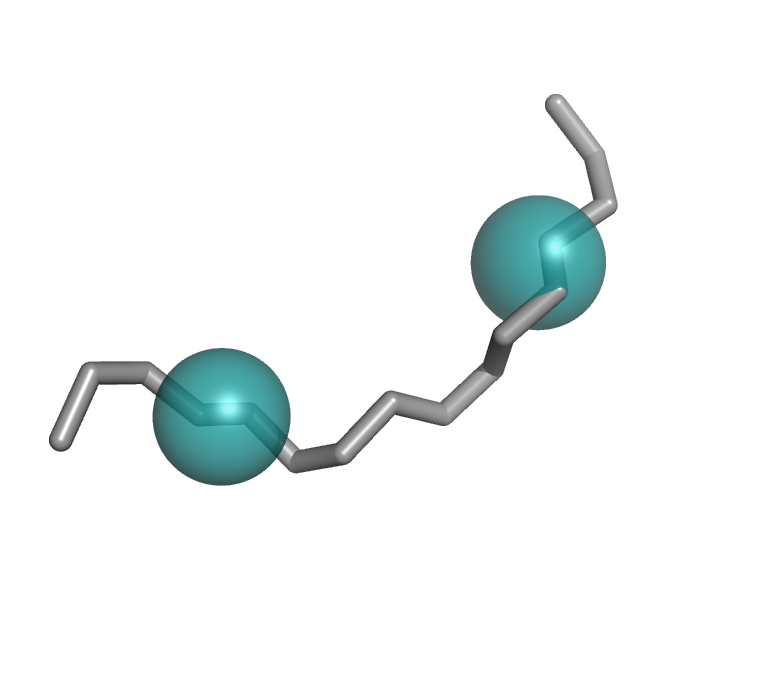}
\caption{\label{fig:2cg} Hybrid coarse-grained/atomistic representation of hexadecane. Only intramolecular forces (bonds, angles, dihedrals) are calculated between the atoms (gray sticks). Nonbonded interactions are only calculated between the coarse-grained interaction sites (cyan spheres) and the resulting forces are distributed between the atoms according to their mass.}
\end{figure}

To avoid this, we have introduced a hybrid model combining a full atomistic representation of bonded interactions with a coarse-grained representation of non-bonded interactions (Figure \ref{fig:2cg}). The alkane chain was divided into subchains (in the cases covered here, two), each of which is represented as a coarse-grained interaction site located at its center of mass. In the coarse-grained region, non-bonded interactions are only calculated between those coarse-grained interaction sites, while the resulting forces are distributed (according to mass) to the atoms represented by these sites. Bonded interactions on the other hand are only calculated in the atomistic representation. While there is no explicit representation of bonds between the coarse-grained interaction sites, the fact that they are always located at the center of mass of the atoms they represent causes them to behave like bonded particles due to the bonded interactions and movement of the underlying atomistic representation. 

The coarse-grained interaction sites are treated exactly the same way as those representing whole molecules in the single-bead AdResS systems. The interaction potentials between the non-bonded interaction sites were derived (as for the one-sphere case) using iterative Boltzmann inversion. To account for the bonded interactions resulting from the underlying atomistic representation, a weak harmonic bond (calculated by noniterative Boltzmann inversion) was added in the iterative Boltzmann inversion calculations, which were not included in the AdResS setups.

\begin{figure}
\includegraphics[width=0.9\textwidth]{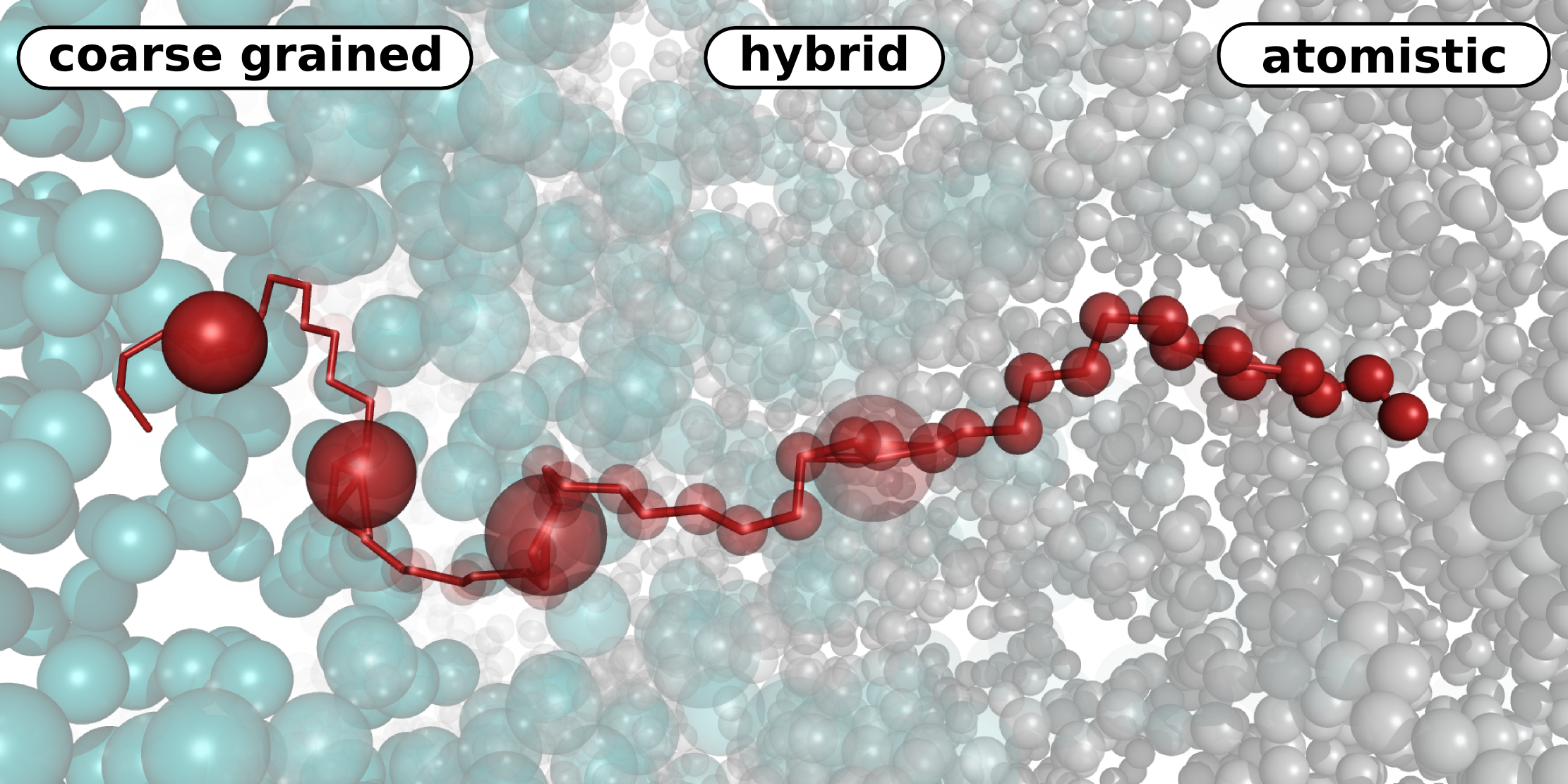}
\caption{\label{fig:pentacontane} Pentacontane (50 monomers) across the hybrid region. Small spheres represent atomistic interaction centers, big spheres represent coarse-grained interaction centers. Bonded interactions (represented by lines in the highlighted molecule) are calculated between atoms in all regions of the AdResS system.}
\end{figure}

\begin{figure}
\includegraphics[width=0.9\textwidth]{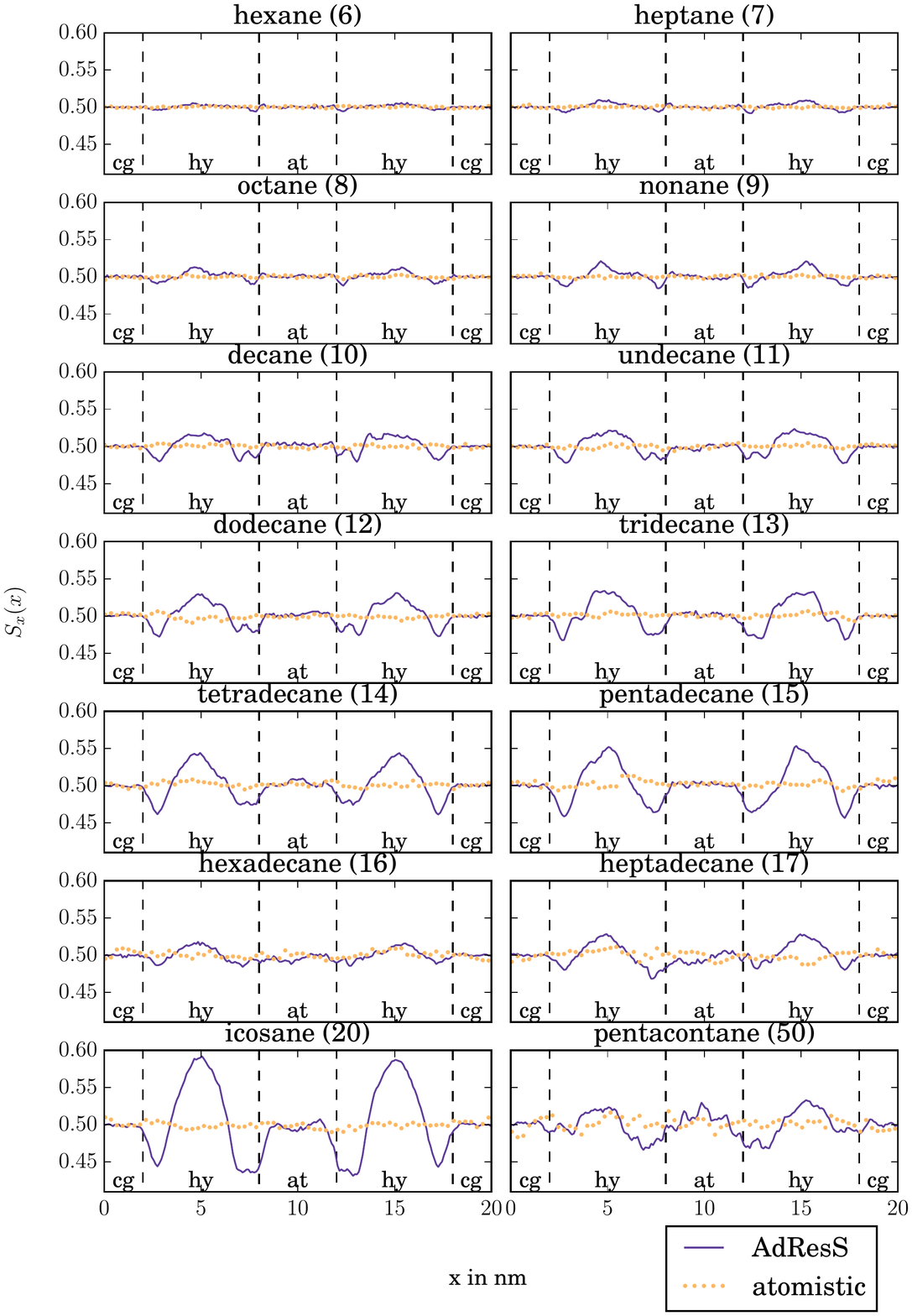}
\caption{\label{fig:OrderParameter} Order parameter (x component) along the x axis of the simulation box, averaged over 9000 frames from a 90 ns simulation after a 10 ns equilibration period. Result from AdResS (blue lines) and full atomistic simulation (yellow dots). The hybrid regions of the AdResS box are marked in gray, the atomistic region is located in the middle, the coarse-grained region at both ends of the plot.}
\end{figure}

To show that this approach can be used for even longer polymers, we have included a simulation of pentacontane, where 50 monomers are represented by 5 interaction sites in the coarse-grained model, which interact through the same interaction potential already used for the two interaction sites of icosane. At this length, extended chains can span the whole hybrid region (Figure \ref{fig:pentacontane}), with one end interacting mainly in the coarse-grained representation and the other in the atomistic representation. Adaptive resolution simulations of this significantly longer polymer correspond to fully atomistic simulations as well as those of shorter chains  (or in some cases even better as for the order parameter, see Figure \ref{fig:OrderParameter}, compared to the less than half as long icosane chain).

By maintaining a representation of the intramolecular structure of the molecule, we avoid the non-trivial problem of reconstructing proper bond-lengths and angles when a molecule crosses from the coarse-grained to the hybrid region, while still achieving a significant reduction of the computational complexity of the simulation. 

From a methodological perspective, this multibead representation may be employed as a first (morphological) prototype to represent large molecules characterized by chemical units with marked differences in reactivity, e.g., hydrophobic and hydrophilic groups. A similar hybrid representation has already been successfully used in simulations of proteins using the MARTINI coarse-grained force field \cite{RzepielaEtAl2011,WassenaarEtAl2013}, making this a promising approach to extend AdResS to even more complex systems.

\section{Soft-Core Potentials in AdResS}
\label{sec:sc}
\begin{figure}
\includegraphics[width=0.9\textwidth]{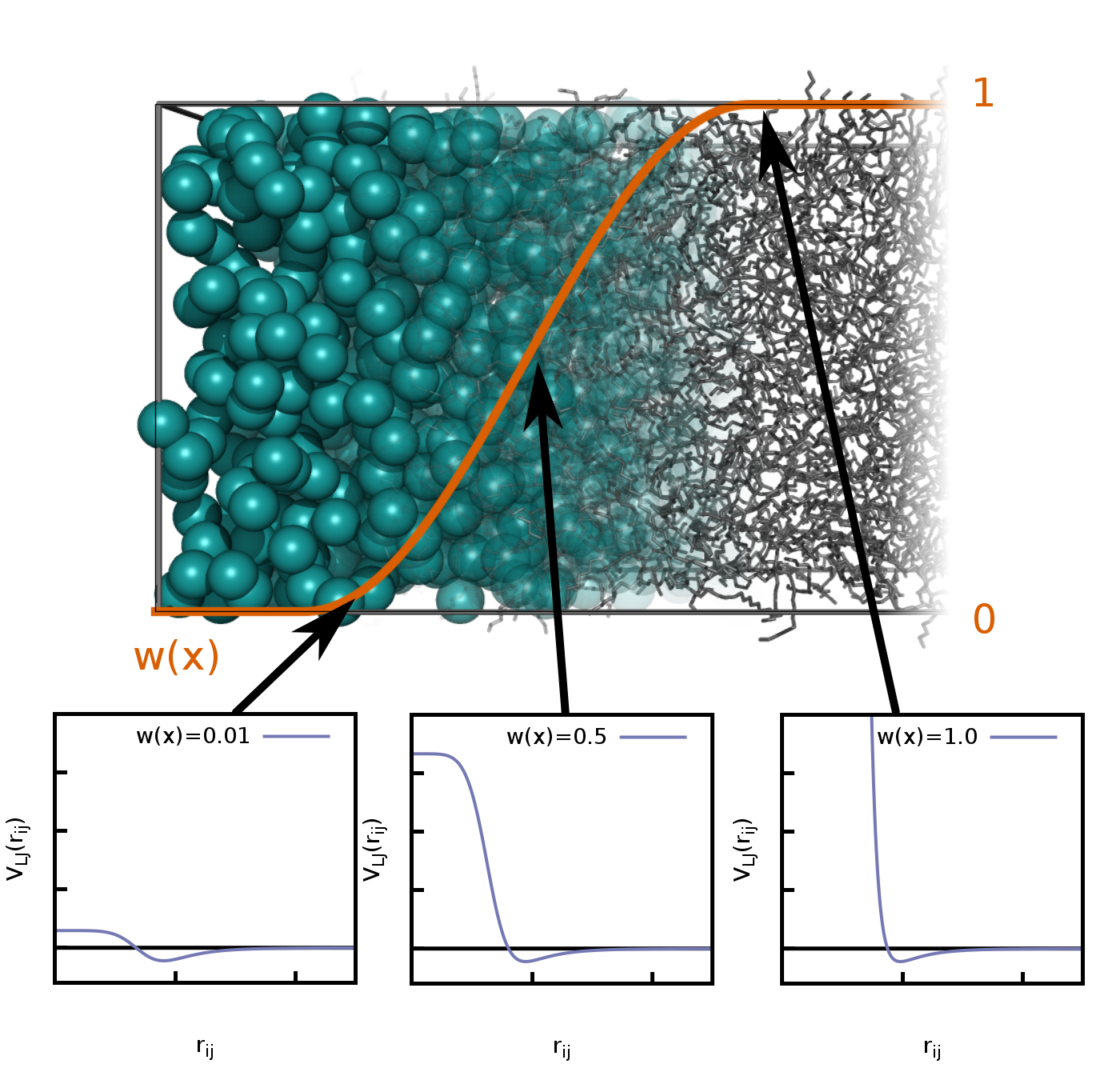}
\caption{\label{fig:scpot_system} Illustration of the use of the soft-core potential in the AdResS system. Near the coarse-grained region, the effect of the soft-core potential is the strongest to prevent numerical instabilities due to clashes between atoms that are entering the hybrid region. Near the atomistic region, the soft-core potential is almost identical to the standard Lennard-Jones potential. Here, any clashes that occur when the molecule enters the hybrid region should have been relaxed.}
\end{figure}

In the hybrid region of an adaptive resolution simulation, molecules are simultaneously represented atomistically and as coarse-grained interaction sites. When a molecule enters into the hybrid region from the coarse-grained region, the atoms and their interactions are reintroduced into the system. As underlined before, in previous applications of AdResS, the molecules treated in this way were small or of an approximately spherical geometry. In these cases, the initial distances between atoms upon entering the hybrid region are ensured due to the repulsive interaction of the coarse-grained interaction sites. For polymer chains, this is not necessarily the case, and a molecule entering the hybrid region can cause atoms to interact at very short distances, resulting in numerical instabilities of the Lennard-Jones potential (Eq. \ref{eq:LennardJones}) used to model atom-atom interactions in the atomistic region.
This problem is similar to problems encountered in alchemical free-energy calculations, in which atoms and their interactions are gradually added to or removed from a simulation system. In order to avoid numerical instabilities in these simulations, soft-core potentials \cite{BeutlerEtAl1994} are employed. Inspired by this similitude, we introduce the following technical modification:\\
The Lennard-Jones potential
\begin{equation}
 \label{eq:LennardJones}
  V_{LJ}(r_{ij}) = \frac{C_{ij}^{(12)}}{r_{ij}^{12}} - \frac{C_{ij}^{(6)}}{r_{ij}^{6}}
\end{equation}
approaches a singularity for $r_{ij} \to 0 $. To avoid the singularity, an effective radius  
\begin{equation}
 r_{\mathrm{eff},ij} = \left( \alpha \sigma^6 w(\mathbf{x})^P + r_{ij}^6 \right) ^ {1/6}
\end{equation}
is used to shift the potential.  Here, $\alpha$ and $P$ are simulation parameters, $\sigma = (C_{12}/C_{6})^{1/6}$ is the radius of the interaction. By including the AdResS weighting function $w(\mathbf{x})$, the potential is gradually changed to the standard Lennard-Jones potential when the molecule moves through the hybrid region into the atomistic region (see Figure \ref{fig:scpot_system}). This is a complementary approach to the force capping method previously used in the treatment of MARTINI water in AdResS\cite{ZavadlavEtAl2014}.

\section{Calculation of structural Properties: Validation of the method}
\label{sec:alkanes}
The validity of the method in simulating liquids of polymer chains is proven in this section. We calculate several structural observables using AdResS and compare them with the results obtained in the equivalent subsystem of a full atomistic simulation. We consider the following quantities: the liquid density $\rho$, the center of mass-center of mass radial distribution function $g_{CoM}(r)$, the atom-atom radial distribution function between two atoms of two different chains $g_{C1-C1}(r)$, the atom-atom intramolecular radial distribution function $g_{intra}(|{\bf r}_{i}-{\bf r}_{j}|)$, the distribution of gyration radius $P(R_{g})$, and the distribution of the end-to-end distance $P(R_{ee})$.

\subsection{Density}
\begin{figure}
\includegraphics[width=0.9\textwidth]{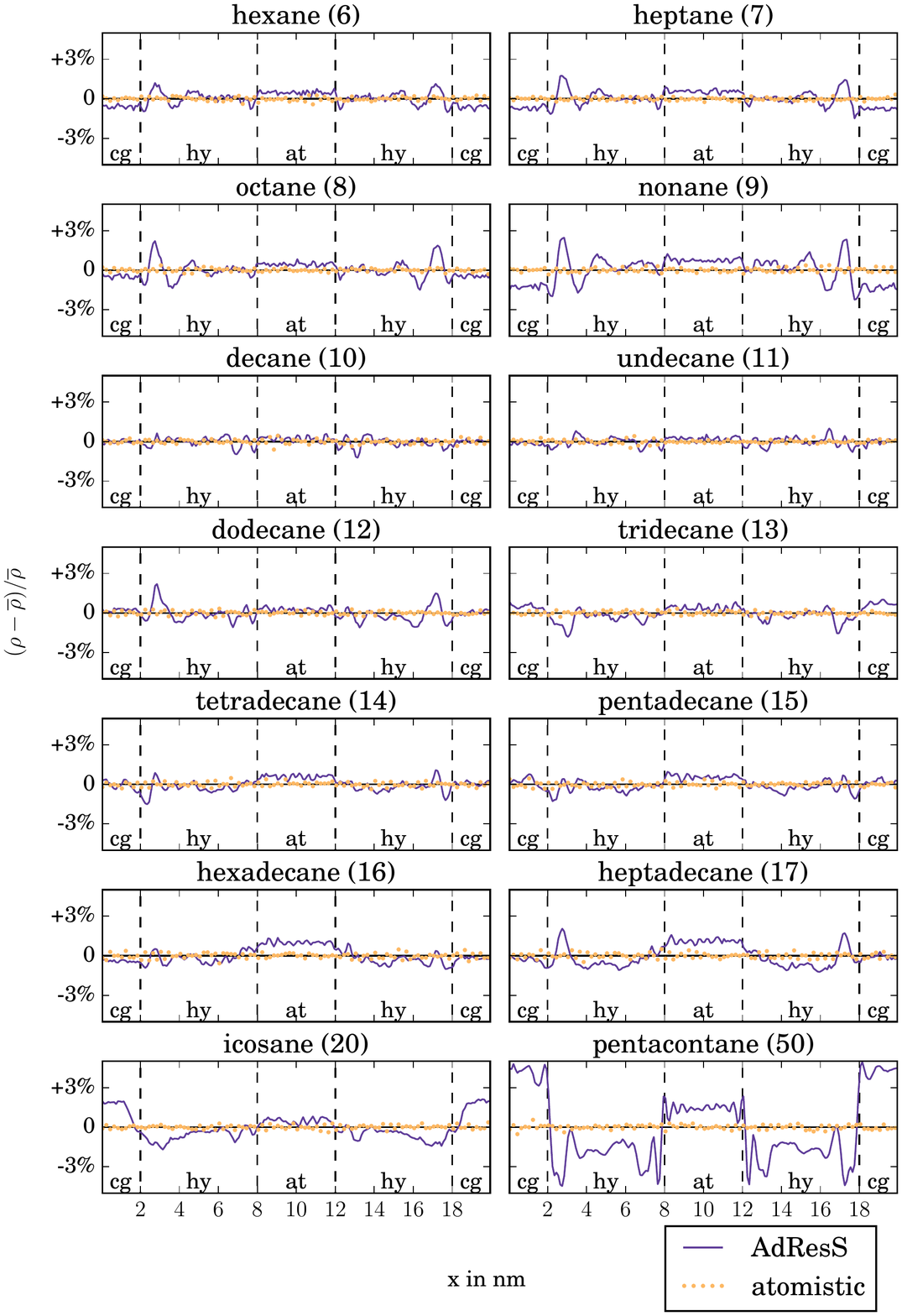}
\caption{\label{fig:density}Density profile along the x axis, averaged over 9000 frames from a 90 ns simulation after a 10 ns equilibration period. Result from AdResS (blue lines) and full atomistic simulation (yellow dots). The hybrid regions of the AdResS box is marked in gray, the atomistic region is located in the middle, the coarse-grained region at both ends of the plot.}
\end{figure}

The capability of AdResS to reproduce the density of the liquid across the box, and in particular in the atomistic region is a mandatory condition for the reliability of the method. A sizable difference in the average density (in the atomistic region) induces automatically a change in the thermodynamic state of the systems which is no more compatible with the equivalent of the full atomistic simulation.\\
In order to calculate the density, the simulation system was divided into 200 slices along the x axis (along which the transition from coarse-grained to atomistic occurs). For each slice, the average mass density of the contained atoms was calculated over the duration of the simulation (Figure \ref{fig:density}).\\
Table \ref{tab:density} reports the maximum difference of density between the atomistic and the coarse-grained region in percentage, as it can be seen the worst case is that of a difference of $2\%$ which is highly satisfactory.
\begin{table}
\caption{\label{tab:density} Difference in density between coarse-grained and explicit region after 20 iterations of thermodynamic force calculation. The resulting thermodynamic force was used for a 100 ns simulation, the average density along the x axis of the system was calculated after the first 10 ns had been ignored for equilibration. All densities in $kg/m^3$.}
\begin{ruledtabular}
\begin{tabular}{cddd}
Molecule&\mbox{explicit}&\mbox{coarse-grained}&\mbox{relative difference}\\
\hline
Hexane      & 632 \pm 1 & 625 \pm 1 & 1.1\% \\
Heptane     & 669 \pm 1 & 661 \pm 1 & 1.3\% \\
Octane      & 687 \pm 1 & 681 \pm 1 & 0.9\% \\
Nonane      & 695 \pm 1 & 680 \pm 1 & 2.0\% \\
Decane      & 711 \pm 1 & 710 \pm 1 & 0.2\% \\
Undecane    & 716 \pm 1 & 714 \pm 1 & 0.3\% \\
Dodecane    & 727 \pm 1 & 726 \pm 1 & 0.2\% \\
Tridecane   & 737 \pm 1 & 739 \pm 2 & 0.2\% \\
Tetradecane & 745 \pm 1 & 741 \pm 2 & 0.6\% \\
Pentadecane & 749 \pm 1 & 746 \pm 2 & 0.4\% \\
Hexadecane\footnotemark[5]  & 759 \pm 1 & 749 \pm 1 & 1.3\%\\
Heptadecane\footnotemark[5] & 763 \pm 1 & 749 \pm 2 & 1.3\%\\
Icosane\footnotemark[5]\footnotemark[6] & 767 \pm 2 & 774 \pm 6 & 0.9\%\\
Pentacontane\footnotemark[7] & 746 \pm 20 & 778 \pm 3 & 4.2\%\\
\end{tabular}
\footnotetext[5]{Using two coarse-grained interaction sites per molecule.}
\footnotetext[6]{For Icosane, 40 thermodynamic force iterations were used.}
\footnotetext[7]{Using five coarse-grained interaction sites per molecule.}
\end{ruledtabular}
\end{table}
Complementary to the table is Figure \ref{fig:density} where the difference in percentage between the expected (full atomistic) average density and the AdResS density is reported as a function of the position in space along the box. The agreement is highly satisfactory and once again with fluctuations below $3\%$ occurring in the hybrid region of some systems. Below we will now move towards more specific structural properties which express more explicitly the underlying microscopic structure. 

\subsection{Radial distribution functions}
We report the results regarding three different radial distribution functions. The first is the center of mass-center of mass radial distribution function $g_{CoM}(r)$. This function represents the molecular structure of the liquid, which is also (by virtue of being the optimization target of the iterative Boltzmann inversion used to derive the coarse-grained potential) reproduced in the coarse-grained region. As in the case of the density the reproduction of this quantity in the atomistic region of AdResS is a mandatory condition for the validity of the method, and Figure \ref{fig:rdfCOM} shows a highly satisfactory agreement between the AdResS results and the full atomistic simulation (the curves actually overlap). 

\begin{figure}
\includegraphics[width=0.9\textwidth]{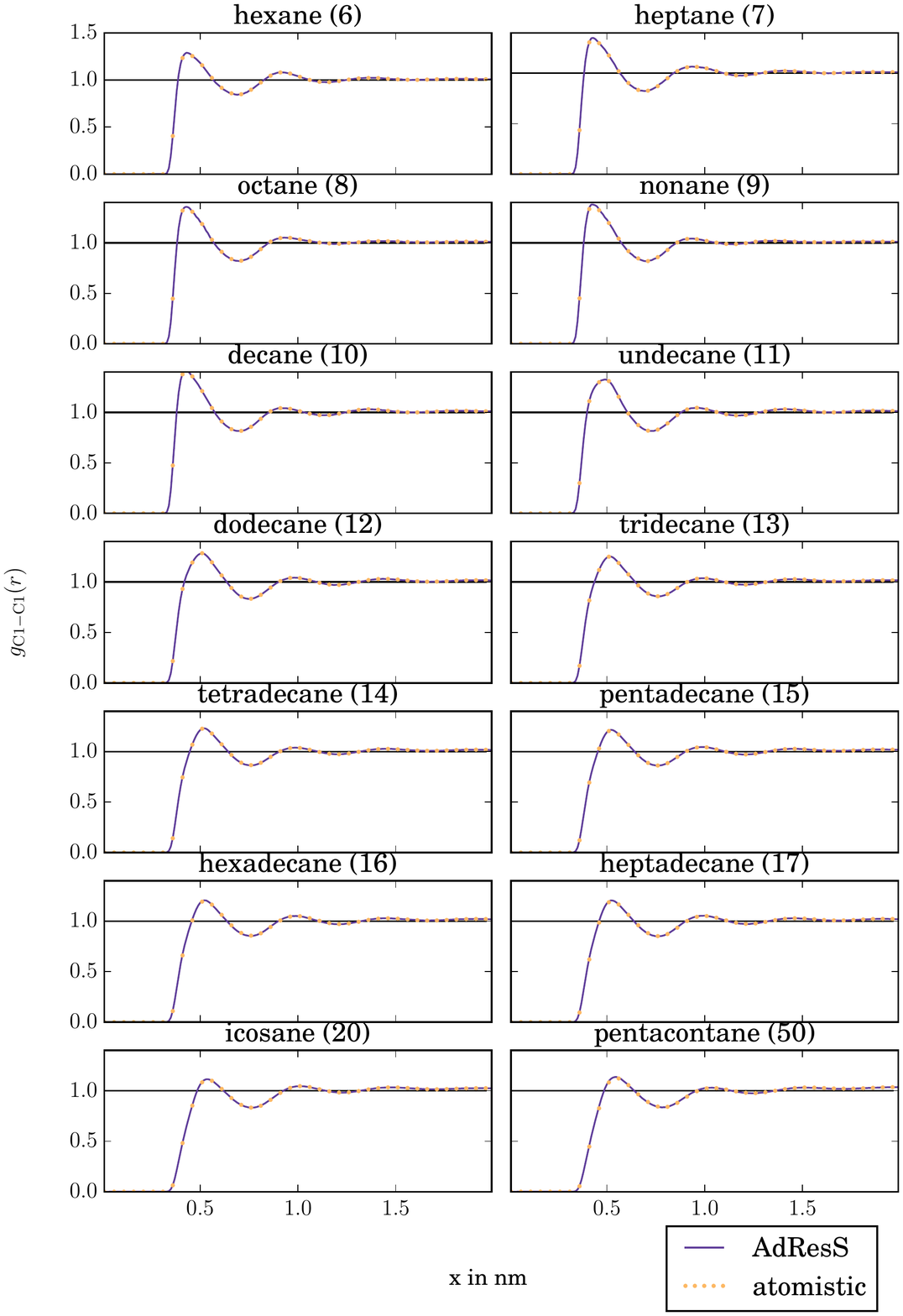}
\caption{\label{fig:rdfC1} Radial distribution function for atom "C1" (the first atom in the polymer chain), calculated for all molecules in the atomistic region over 9000 frames from a 90 ns simulation after a 10 ns equilibration period. Result from AdResS (blue lines) and full atomistic simulation (yellow dots).}
\end{figure}

The second quantity is the interchain atom-atom radial distribution function $g_{CC}(r)$, which is calculated by considering an atom of one chain as a reference with  atoms of the other chains. Here for simplicity we report the specific case of the  function $g_{C1-C1}(r)$ that is the radial distribution function which considers the first carbon of each chain. Figure \ref{fig:rdfC1} shows a remarkable agreement between the AdResS results and the full atomistic simulation; this is particularly important since such a quantity reports a purely atomistic feature of the liquid in contrast to the density and $g_{CoM}(r)$, which are equally represented independently from the molecular resolution (and indeed these two properties are mostly conserved in the coarse-grained region as well). 

\begin{figure}
\includegraphics[width=0.9\textwidth]{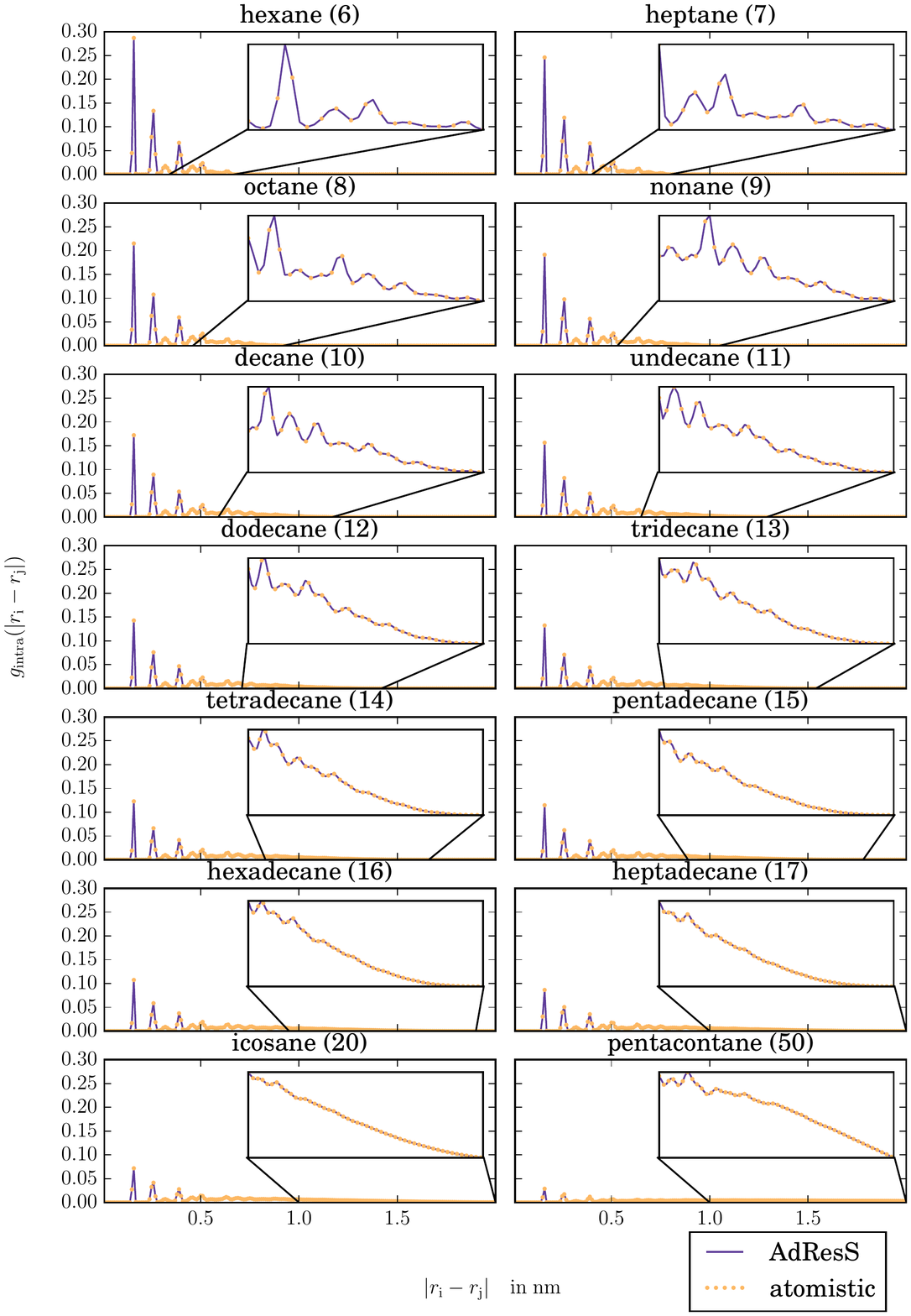}
\caption{\label{fig:idd} Intramolecular atomic radial distribution function, averaged over all molecules in the atomistic region over 9000 frames from a 90 ns simulation after a 10 ns equilibration period. Result from AdResS (blue lines) and full atomistic simulation (yellow dots). As the first peaks of the distribution are determined by the bond-length, the "long-range" tail region is shown in more detail in the insets of the corresponding plots.}
\end{figure}

If in the atomistic region of AdResS, the molecular structure function and the intramolecular structure function are the same as in the full atomistic simulation, then we can be confident that a microscopic analysis of local properties done in AdResS is equivalent to that of a full atomistic simulation. To show this feature, we have calculated in addition the atom-atom intramolecular radial correlation function $g_{intra}(|{\bf r}_{i}-{\bf r}_{j}|)$, where ${\bf r}_{i}$ and ${\bf r}_{j}$ are the positions of two carbons in the same chain. As for the other quantities the agreement is highly satisfactory (see Figure \ref{fig:idd}).\\
A further confirmation of the reliability of AdResS in describing in a fully satisfactory way the structure of even single chains in the liquid is given by the distribution of the radius of gyration $R_{g}$ and by the distribution of the end-to-end distance $R_{ee}$ reported in the next section.

\subsection{Distribution of the Gyration Radius and of the End to End Distance}
The radius of gyration of a polymer consisting of M monomers is defined as

\begin{equation}
R_g = \frac{1}{M} \sum_i^M \left( \mathbf{r_i}-\mathbf{\overline{r}} \right)^2
\end{equation}

\begin{figure}
\includegraphics[width=0.9\textwidth]{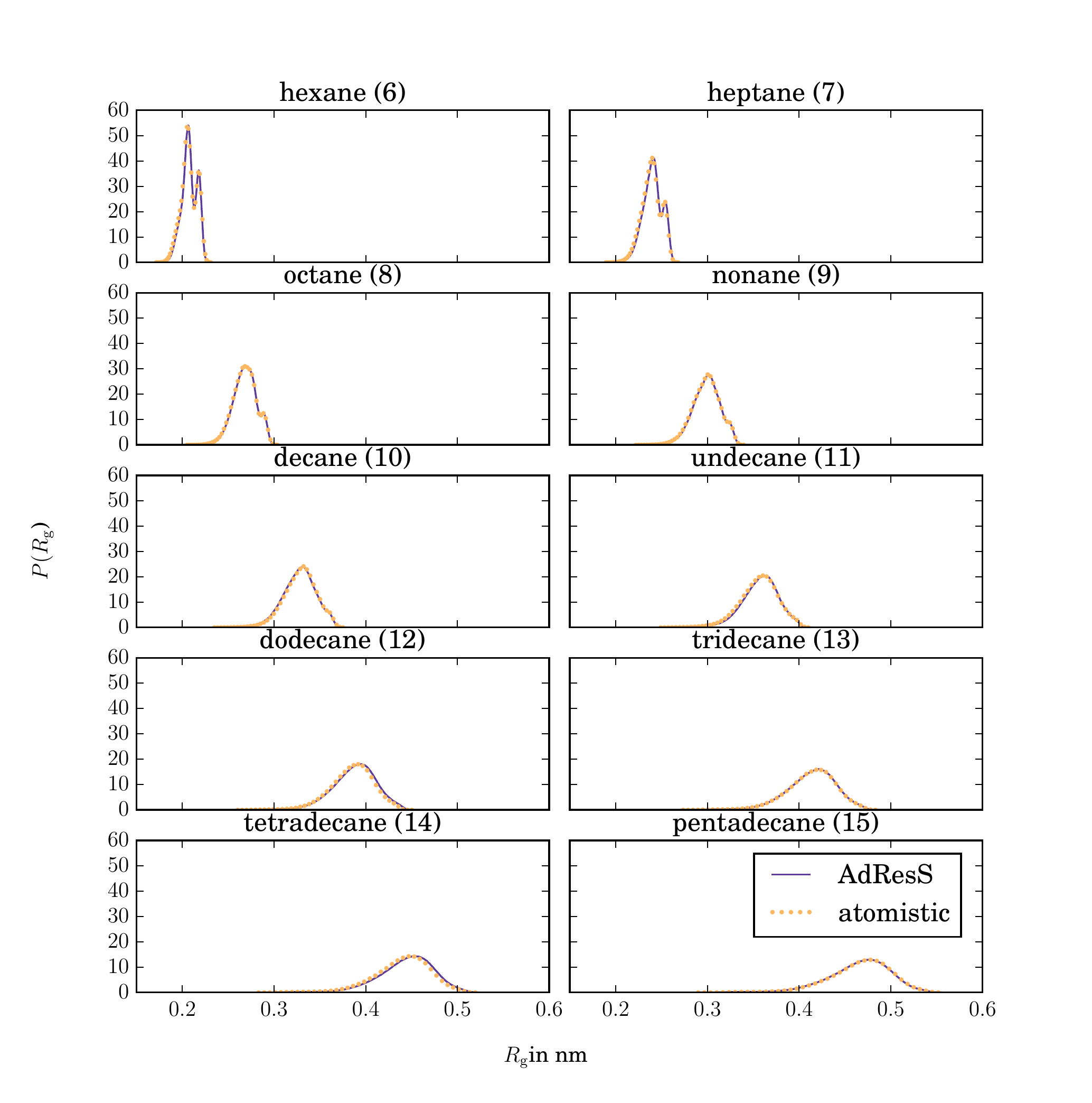}\\
\includegraphics[width=0.9\textwidth]{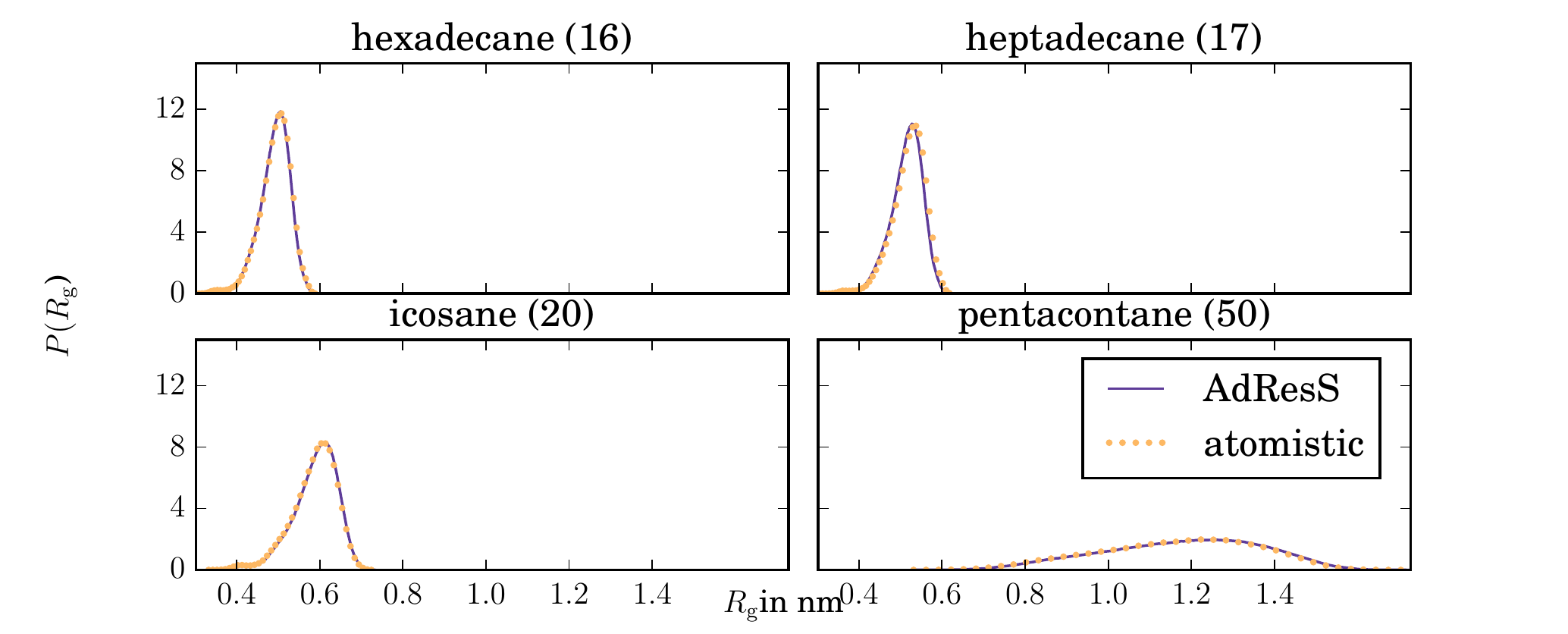}
\caption{\label{fig:PRGyration} Distribution of the radius of gyration,
 averaged over all molecules in the atomistic region over 9000 frames from a 90 ns simulation after a 10 ns equilibration period. Result from AdResS (blue lines) and full atomistic simulation (yellow dots).}
\end{figure}

Where $\mathbf{\overline {r}}$ is the center of mass (geometry) of the molecule. We have calculated the statistical distribution of $R_{g}$ , ($P(R_{g})$), of  all molecules in the atomistic region of AdResS over the duration of the simulation and compared it with the equivalent in a full atomistic simulation.  Figure \ref{fig:PRGyration} reports the results which once again are highly satisfactory. The same calculation was done for statistical distribution of the end-to-end distance, $P(R_{ee})$, where $R_{ee}$ is defined as usual, as
\begin{equation}
R_\mathrm{ee} = | \mathbf{r_1} - \mathbf{r_M} |
\end{equation}
that is, the distance between the two atoms representing the two ends of the polymer chain of $M$ monomers. Once again the agreement is highly satisfactory as illustrated in Figure \ref{fig:PRendend}.

\begin{figure}
\includegraphics[width=0.9\textwidth]{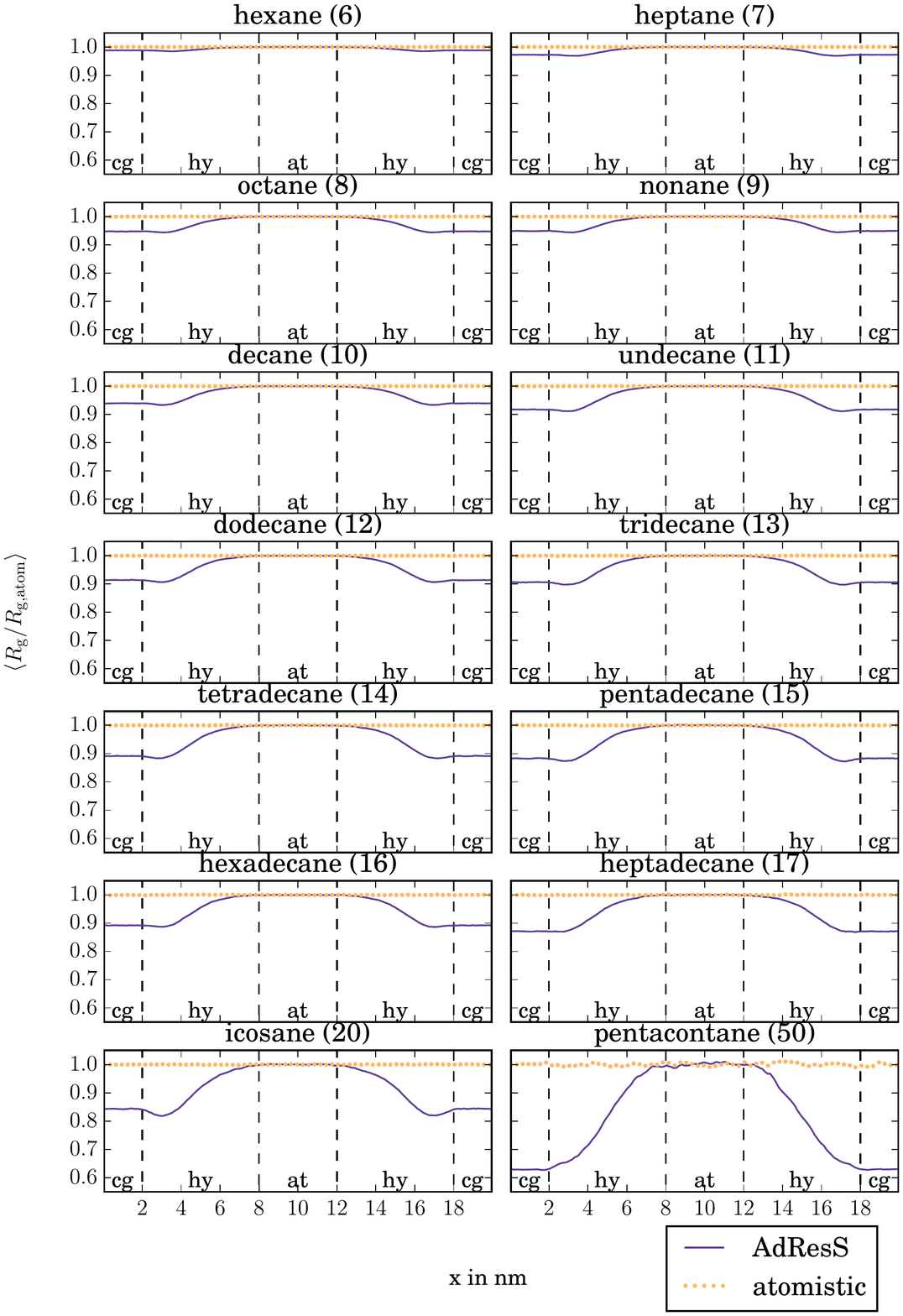}
\caption{\label{fig:LRGyration} Local average radius of gyration along the x axis of the simulation box. Due to the hybrid topology used in the coarse-grained region, the radius of gyration is defined in this region as well. Values are averaged over all molecules in the atomistic region over 9000 frames from a 90 ns simulation after a 10 ns equilibration period. Result from AdResS (blue lines) and full atomistic simulation (yellow dots).}
\end{figure}

The removal of non-bonded atomistic interactions that occurs when a molecule crosses the hybrid region towards the coarse-grained region does not only affect interactions between atoms of different molecules, but also intramolecular non-bonded interactions. As these are generally repulsive, their removal causes the polymer chains to collapse (by up to $40\%$ of their volume, as indicated by the decrease of the radius of gyration in this area depicted in Figure \ref{fig:LRGyration}). While transitioning the hybrid region towards the atomistic region, the polymer chains can increase to the values observed in full atomistic simulations.

\section{Validation of global AdResS behavior}
The previous observations show that in the atomistic region, the behavior of the AdResS system is equivalent to that of a full atomistic simulation. It would however be possible that this would be the case even if the AdResS system would not behave as expected outside of the atomistic region. Two such failure scenarios have been explored in previous studies of AdResS, both related to the influence of the (essentially non-physical) hybrid region on the behavior of the whole system.

First, it has been hypothesized \cite{pre2} that the hybrid region could constitute a kinetic barrier between the coarse-grained and atomistic regions, preventing the exchange of particles. Second, the introduction of an interface into the system could have an ordering effect on the orientation of molecules that are not rotationally symmetric (for example the dipole moment in liquid water \cite{jpcm,jcpwater}). While these concerns have shown to be unfounded in the systems studied before, we have tested our alkane simulations for these effects as well. 

\begin{figure}
\includegraphics[width=0.9\textwidth]{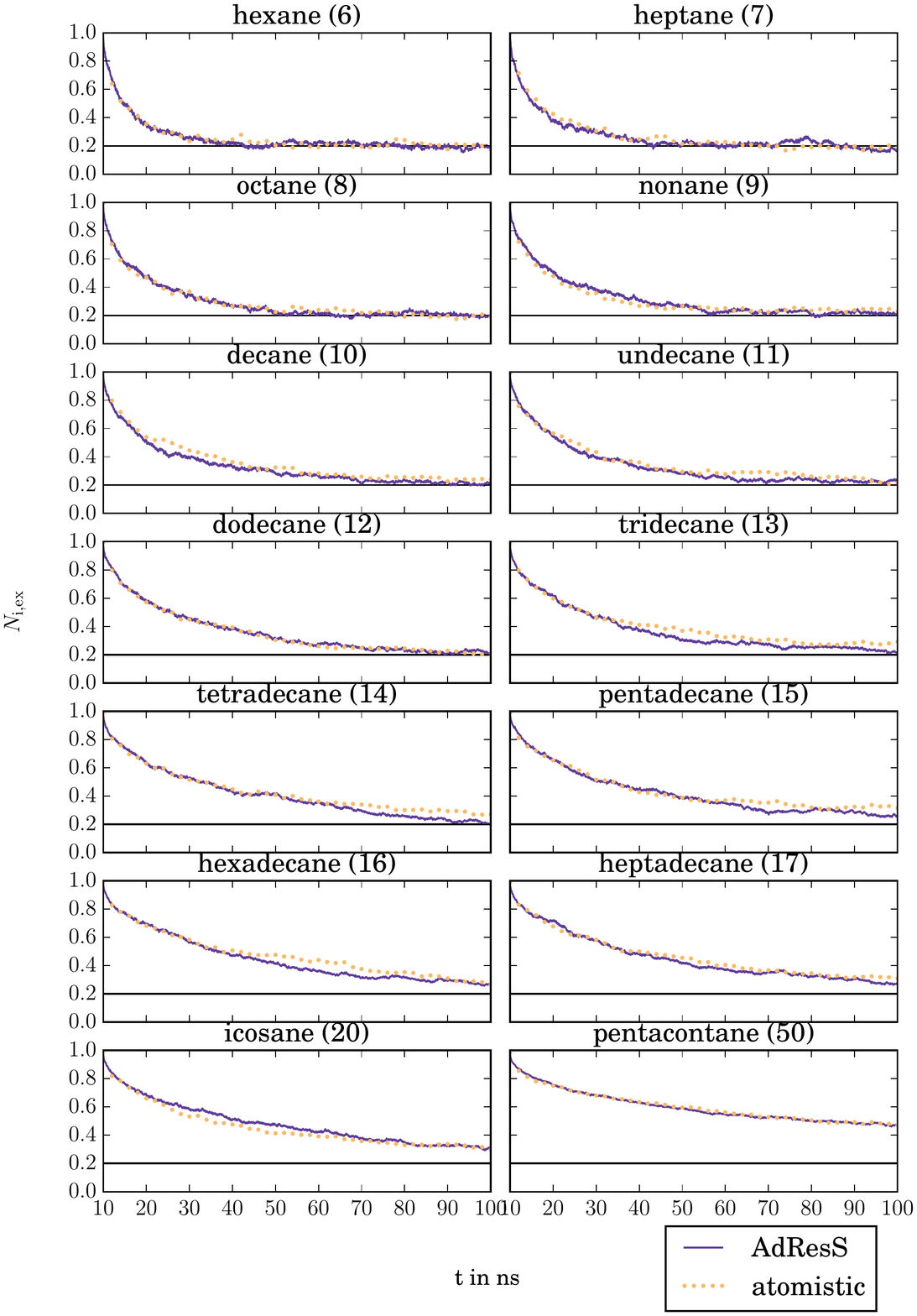}
\caption{\label{fig:drift} Fraction of atoms initially present in the atomistic region and still present in the atomistic region over time. The calculation was done over 90 ns simulation after a 10 ns equilibration period. As the simulation box is finite, the value converges to an equal distribution of all initially present atoms in the box, resulting in a value of 0.2 (the volume-fraction of the atomistic region in the simulation box). Result from AdResS (blue lines) and full atomistic simulation (yellow dots).}
\end{figure}

Figure \ref{fig:drift} shows the fraction of atoms initially present in the atomistic region that are still in this region over time. It can be seen that the exchange of molecules between the atomistic region and the rest of the simulation box is essentially the same in AdResS and in a full atomistic simulation. In an infinite system, the molecules initially present in the atomistic region would diffuse away so that after a long enough time none would be left. In the finite system (with periodic boundary conditions) considered here, these molecules distribute evenly in the box so the fraction of molecules that can be observed in the original volume approaches the volume fraction of this volume (in this case 0.2).

\begin{figure}
\includegraphics[width=0.9\textwidth]{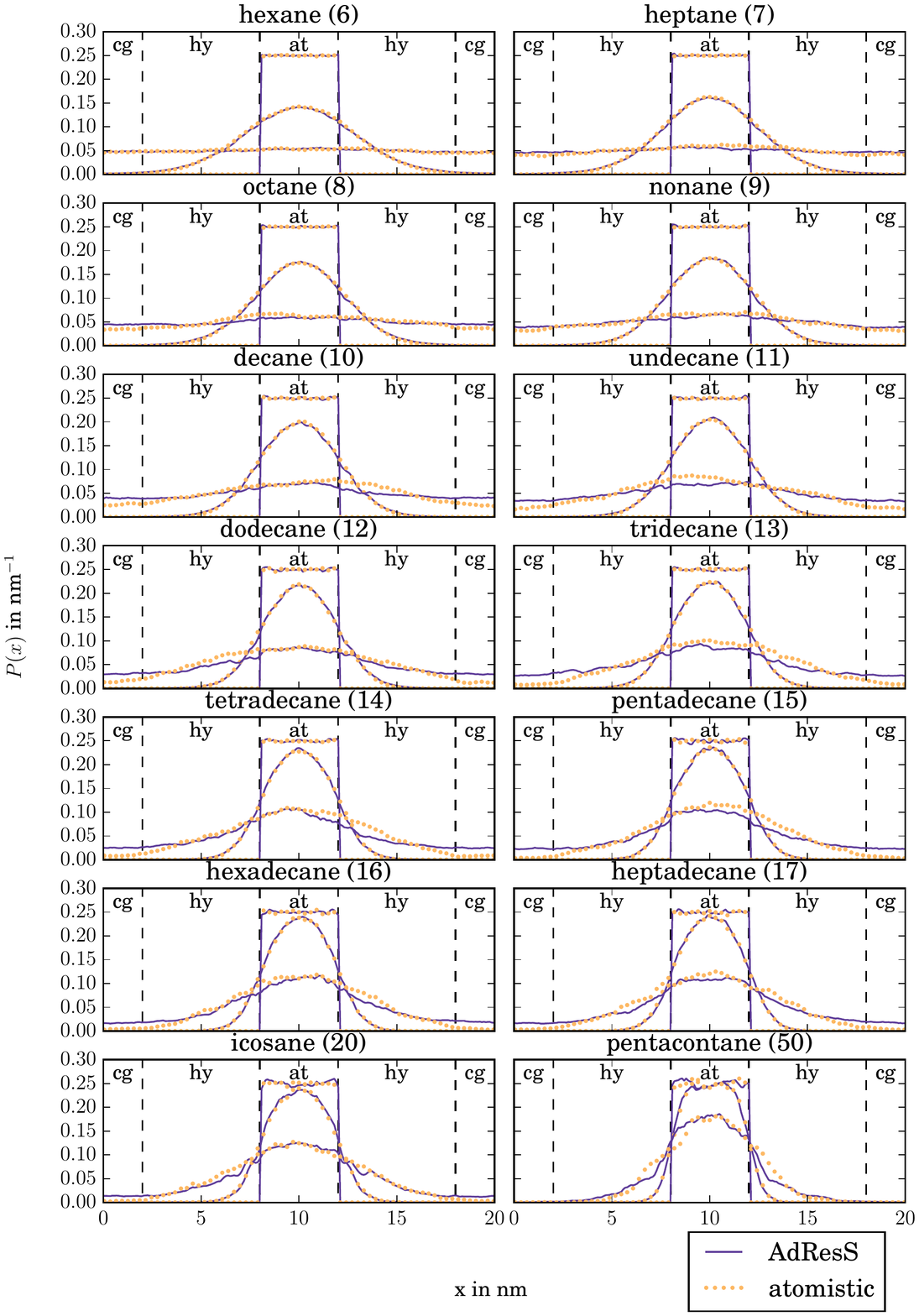}
\caption{\label{fig:diffusion} Position probability density (along the x axis) for a molecule $0$ ns, $4$ ns and $40$ nm after it is present in the atomistic region. For longer chains, the difference between fully atomistic and AdResS profiles in the coarse-grained part of the simulation box is caused by the faster diffusion in the coarse-grained part.}
\end{figure}

Similarly, Figure \ref{fig:diffusion} illustrates the diffusion of particles out of the atomistic region. For this, the observed  probability density for a molecule in the atomistic region after different time periods was calculated. Due to faster diffusion observed in coarse-grained models, these profiles can be expected to differ between AdResS and fully atomistic simulations in the hybrid and coarse-grained regions, while they remain the same in the atomistic region up to statistical fluctuations.

To investigate the possibility of orientational alignment due to the interfaces introduced by the hybrid region, we calculate an indirect orientation parameter along the x axis of the system:
\begin{equation}
S_x = \frac{| (\mathbf{r_1} - \mathbf{r_M}) \cdot \mathbf{x} |}{R_\mathrm{ee}}
\end{equation}
The use of the absolute value in this case is necessary, as an alkane chain has no intrinsic directionality (i.e. defining one end as first monomer and the other as last is completely arbitrary), so that a directed orientation parameter would always have an expectation value of $0$. The expected average value of the indirect $S_x$ for an ensemble of randomly oriented molecules is 0.5. We have calculated the observed value for 200 slices along the x axis of the system averaged over 90 ns of simulation time. Figure \ref{fig:OrderParameter} reports the value of $S_x$ as a function of the position in the box. Although, inevitably, for longer chains there is an interface effect (the maximum overall difference is about $10\%$), at the edge of the atomistic region, in general the result in the atomistic region is very satisfactory. 
One must also consider that the computational set up was built as a ``worst-case-scenario''; that is, we consider a very small atomistic region. In fact if the method works well  enough with such setup, then it will work certainly well for systems with larger atomistic regions. In summary, the results of this section confirm that the actual functioning of AdResS is correct and thus the results of the previous section are indeed a proof of the reliability of AdResS in reproducing full atomistic results.

\section{Computational Performance}
\label{sec:performance}
There are two main features that AdResS offers compared to full atomistic simulations. One, independently from the computational costs, is the possibility of employing AdResS as an analysis tool in order to unambiguously identify the essential (atomistic) degrees of freedom required, in order to obtain a certain property (see, e.g., Ref.\cite{jcpfuller}). This can result in a better understanding of the process and eventually enable its manipulation in order to trigger properties on demand. The second is the lowering of the computational costs associated with the drastic reduction of the number of degrees of freedom. 

To realize this gain, the system has to be large enough to offset the additional computational complexity associated with the introduction of the hybrid region, where not only both the coarse-grained and atomistic interaction have to be calculated, but also both of these interactions have to be interpolated, resulting in additional calculations. 

\begin{figure}
\includegraphics[width=0.9\textwidth]{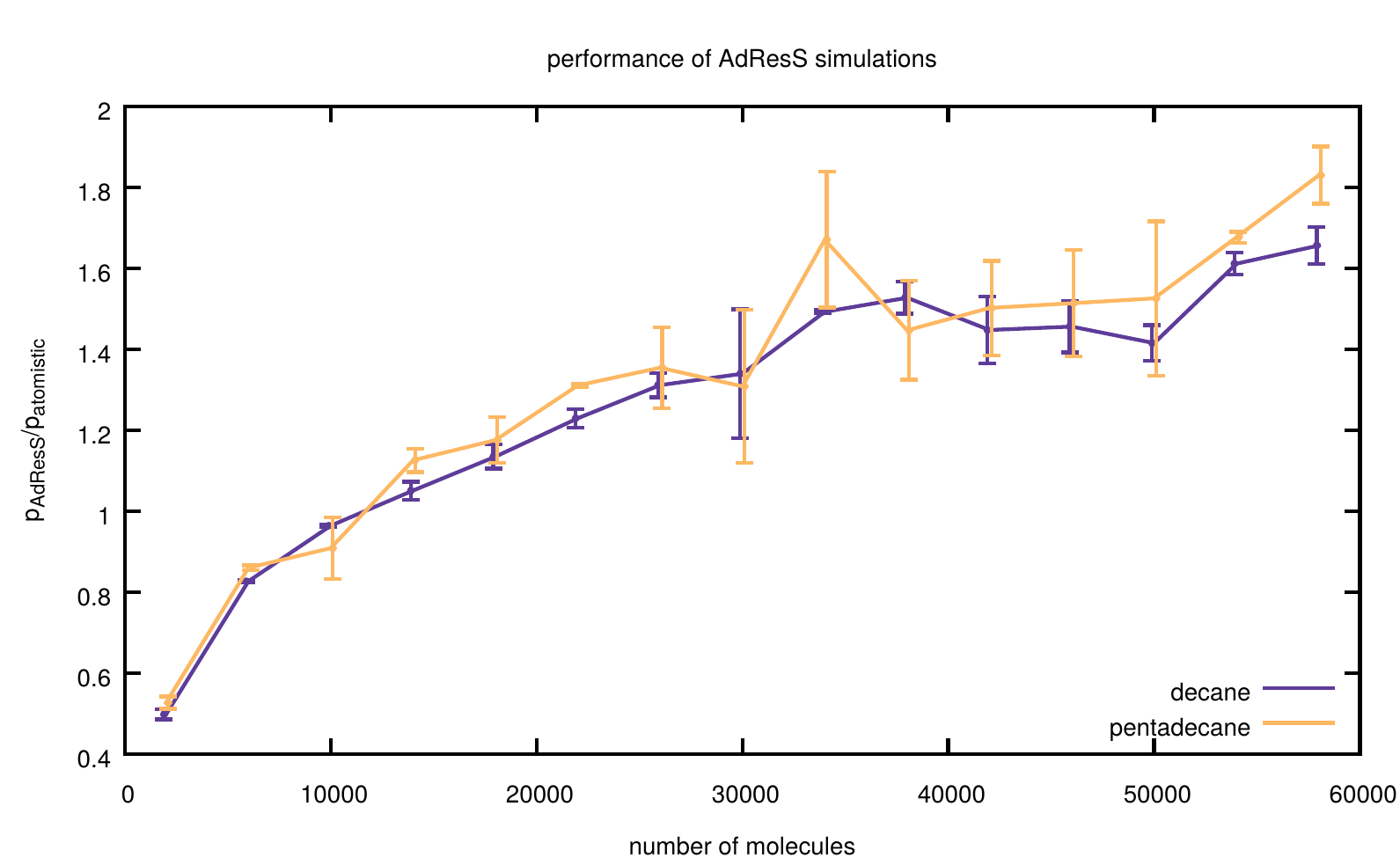}
\caption{\label{fig:performance} Estimation of the performance of AdResS relative to a full atomistic simulation depending on the total system size (number of molecules in the simulation). For each system size, five short (5000 steps) simulations were conducted of both the full atomistic simulation and AdResS system under production conditions (on 2x Xeon Westmere X5650 hexacore processors). Performance according to GROMACS logfile (in ns of simulation time per day of walltime). Variations (estimated from standard deviations between each set of five simulations) can be expected in such short simulations due to GROMACS load balancing, domain decomposition, and other random factors. The size of the atomistic and hybrid regions remained constant (the atomistic region contains on average 400 molecules). }
\end{figure}

As this work is concerned with the development and validation of the method, the geometries of the systems studied here were not chosen for computational efficiency and actually must be considered as a ``worst case scenario'' technical set up. Indeed, as the hybrid region actually makes up the biggest part of the simulation boxes, the performance was necessarily worse than that of a full atomistic simulation.
However, some applications require a significantly larger total system size (see, eg. Ref.\cite{schulten} and references therein) and with an increased size of the coarse-grained region, the computational efficiency of AdResS in comparison to full atomistic simulations increases.  Figure \ref{fig:performance} shows the performance of AdResS compared to a full atomistic situation with increasing system size. Interestingly for system sizes of the order of large material science and biophysical simulations by now routinely treated in literature, the computational gain is, while not spectacular, already sizable. It should be noted that the performance gains approach the value observed for a system in purely coarse-grained representation in the current implementation ($\approx 2$).

However, the plot actually shows a lower bound of the performance, which could be improved in several ways, such as introduction of multiple time steps, optimization of the size of the hybrid region and a switch to spherical symmetry. To utilize all such computational improvements one must add a code optimization within the particular package employed. The aim of this work is to give a proof of concept that AdResS can correctly describe the structural properties of complex systems made of large complex molecules; however already at this stage the computational gain is sizable, for systems large in size but of routine interest in current research.

\section{Changing the Size of the Hybrid Region}

\begin{figure}
\includegraphics[width=0.9\textwidth]{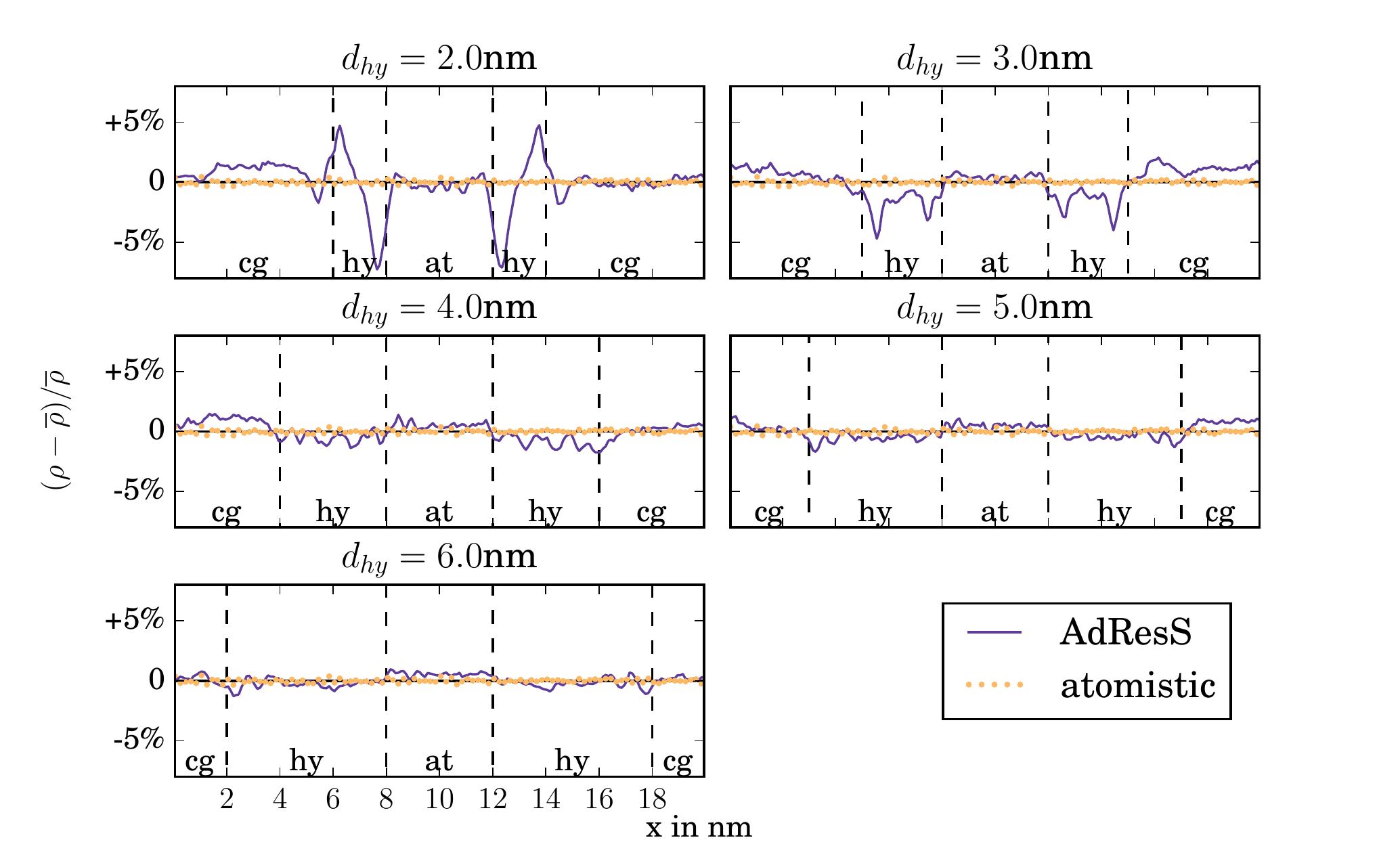}
\caption{\label{fig:DensityHy} Density profile along the x axis (as in Figure \ref{fig:density}) for pentadecane using different widths of the hybrid region.}
\end{figure}

\begin{figure}
\includegraphics[width=0.9\textwidth]{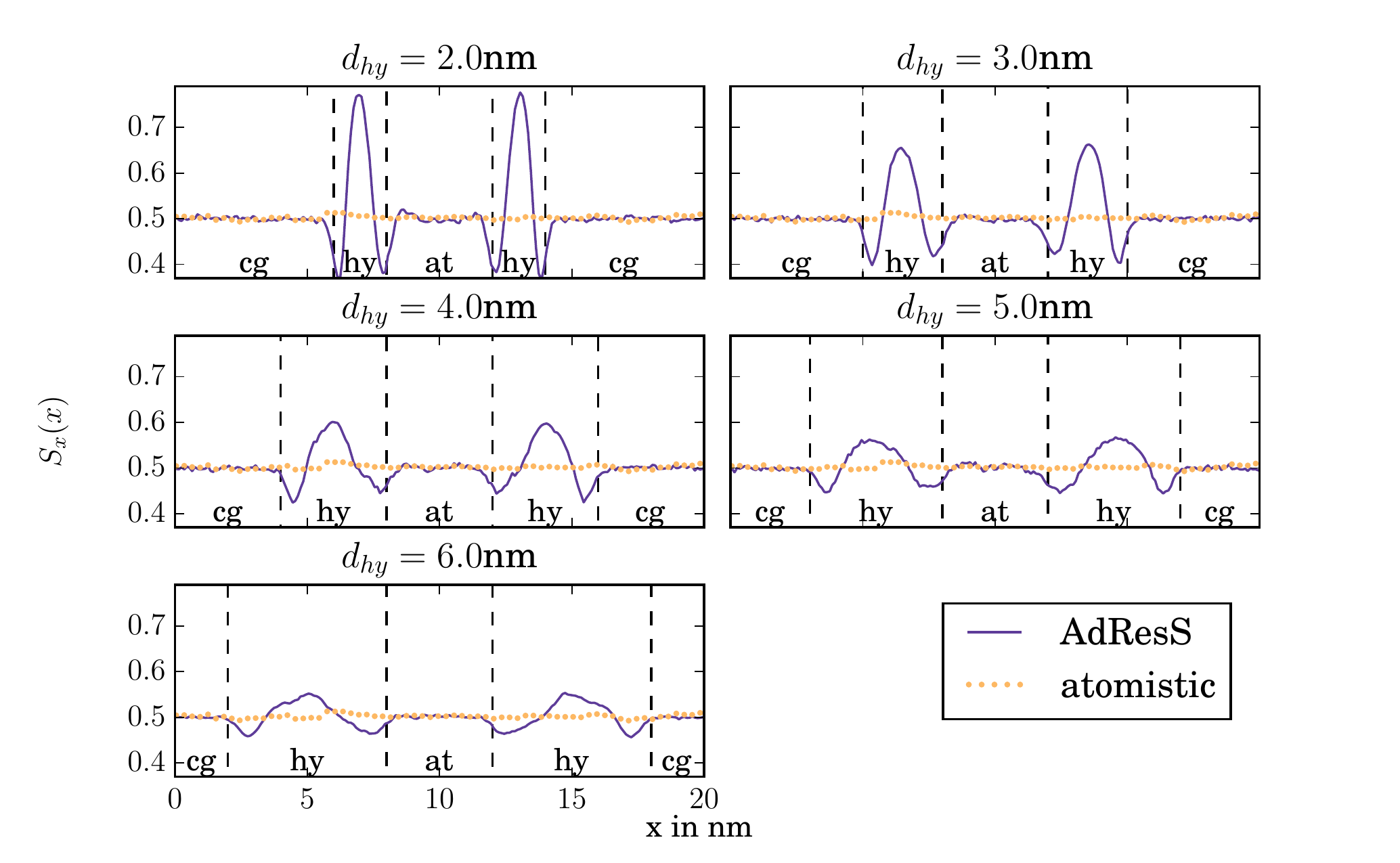}
\caption{\label{fig:OrderparameterHy} Order parameter (x-component) along the x axis of the simulation box (as in Figure \ref{fig:OrderParameter}) for pentadecane using different widths of the hybrid region.}
\end{figure}

\begin{figure}
\includegraphics[width=0.9\textwidth]{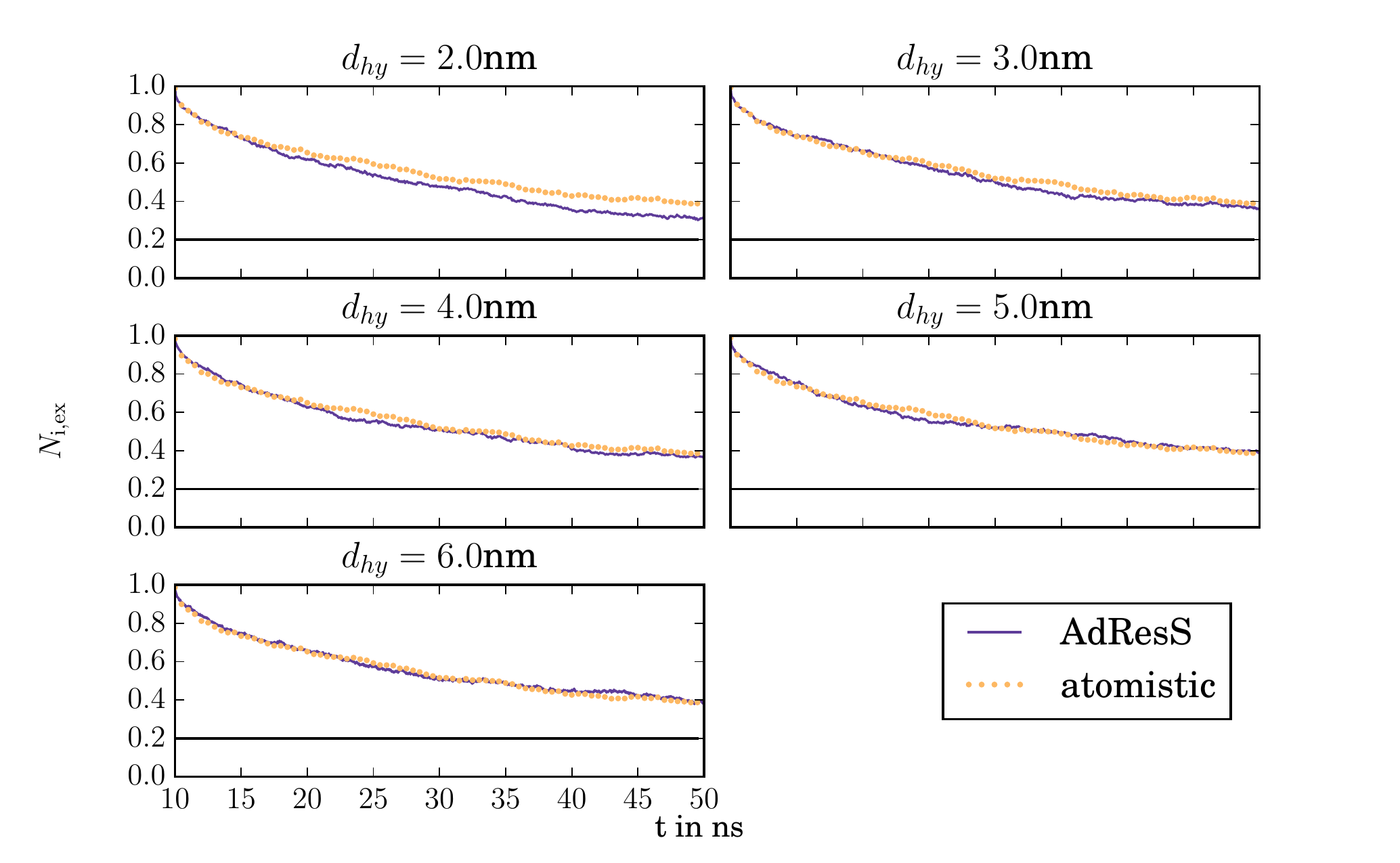}
\caption{\label{fig:DriftHy} Fraction of atoms initially present in the atomistic region and still present in the atomistic region over time (as in Figure \ref{fig:drift}) for pentadecane using different widths of the hybrid region.}
\end{figure}

As mentioned in Sec. \ref{sec:performance}, the size of the hybrid region strongly influences the total performance of an adaptive resolution simulation, as both atomistic and coarse-grained interactions have to be computed in this region. Hence, from a performance point of view it is desirable to reduce the size of the hybrid region as much as possible. In the simulations presented up until now, a relatively large hybrid region ($d_{hy}=6.0$nm) has been chosen to ensure computational stability. Using pentadecane (the system with the highest number of atoms per coarse-grained interaction site in this study) as an example, we have tested the influence of different sizes of the hybrid region on the resulting simulation ensemble. We calculated the thermodynamic force using the same protocol as before (described the Appendix) and ran 50 ns simulations for $d_{hy}=2.0/3.0/4.0/5.0$nm. As with $d_{hy}=6.0$nm, structural properties (rdf, intramolecular rdf, radius of gyration, and end-to-end distance) of a fully atomistic simulation were reproduced well in the atomistic region for all sizes of the hybrid region (not shown). Increased fluctuations of density (Figure \ref{fig:DensityHy}) and the order parameter (Figure \ref{fig:OrderparameterHy}) can be observed, especially for the smallest hybrid regions considered ($d_{hy}=2.0$nm and $d_{hy}=3.0$nm), while diffusion out of the hybrid region still occurs as fast as in the atomistic simulation (Figure \ref{fig:DriftHy}).

As with the performance values presented in Sec. \ref{sec:performance}, these results represent the current implementation of AdResS and the thermodynamic force calculation protocol. Future technical improvements of both algorithms are likely to improve the results.

\section{Conclusions}
We have presented a systematic study of liquids of alkane chains of different lengths employing the AdResS scheme modified for large molecules. This study is a proof of concept of the reliability of AdResS for systems containing large polymer-like molecules. This is a technical and conceptually relevant result because although AdResS has been applied in the past to an extended range of interesting systems, the molecules composing these systems did not possess the extended structure of long chains. This implies that for previous systems the change of resolution of a molecule was not as dramatic as it can be for a chain with large fluctuation of its shape in space and with the possibility of entanglement. We have shown that the extension of AdResS done in this work overcomes these difficulties in a rather efficient way and the resulting technique can reproduce with high accuracy the results of full atomistic simulations and thus can be used as a powerful tool for local microscopic analysis of physical properties at lower computational cost compared to full atomistic simulation (at least for large systems, e.g., of biophysical interest). This work opens new perspectives to the application of AdResS to systems that span from polymer mixtures, to biological systems.
\section*{Acknowledgment}
This research has been funded by Deutsche Forschungsgemeinschaft (DFG) through Grant CRC 1114, Project C01. The authors acknowledge the North-German Supercomputing Alliance (HLRN) for providing HPC resources that have contributed to the research results reported in this paper.

\appendix

\section{Parameters and Simulation protocol}
\label{sec:simulation_parameters}
All atomistic simulations were performed using the GROMACS simulation package (version 5) while AdResS results were produced with a modified version of GROMACS release 5.0.6. \cite{Abraham_GROMACS_2015}. For both atomistic simulations and AdResS, the leap-frog stochastic dynamics integrator (sd) was used with force cut-off length of $0.9$nm.

For the explicit representation of the alkane chains, topologies based on the gromos 53a6 force field have been taken from the Automated Topology Builder and Repository \cite{MaldeEtAl2011} with the exception of pentacontane, whose topology was created based on that of icosane. To reduce the computational complexity, a united atom model was used, allowing a 4fs time step to be used in both atomistic simulation and AdResS.

Each alkane system considered was set up as a rectangular simulation box with 2000 molecules with periodic boundary conditions. A 10 ns full atomistic simulation with anisotropic pressure coupling (to keep the long side of the box at a constant 20nm) using a Berendsen barostat ($\tau_p=1$ps) was performed to achieve an equilibrated starting conformation and a density appropriate to the model used (all further simulations were performed at constant volume). As the aim of the study was to investigate liquid systems, simulations of hexane up to heptadecane were performed at 300K, while icosane (with a melting point of approx. 310K) was simulated at 350K (due to the use of stochastic dynamics, no additional thermostat was applied) and pentacontane at 400K. 

To create a starting structure for pentacontane, chains were generated in a box using a biased random walk using bond-length, bond angle and dihedral angle distributions from the structure of an equilibrated icosane melt. After a steepest descent energy minimization and the removal of clashing chains, the system was equilibrated in a 20 ns annealing simulation, where it was heated up  from 400K to 800K and cooled back down to 400K. 

A 100 ns atomistic simulation provided the target distribution for the iterative Boltzman inversion as well as a reference system for comparison to the AdResS results.

The AdResS transition was set up along the long side of the simulation box (fixed to 20nm) with a 4nm wide atomistic region, two 6nm wide hybrid regions and a 4nm wide coarse-grained region across the periodic boundary along the x axis of the system (Figure \ref{fig:system}).

The thermodynamic force was calculate in 20 iterations (with the exception of icosane, where 40 iterations were used) with 1.6 ns simulation time each using the VOTCA toolbox. In the hybrid region a soft-core potential (described in section \ref{sec:sc}) was used for distances up to a cut-off of $0.5$nm.

The thermodynamic force resulting from the last iteration was then used in a 100 ns AdResS run whose properties were compared with those of the 100 ns full atomistic simulation performed before. For this, 9000 conformations (starting at 10 ns simulation time) were used from each simulations.

\section{Figures}
\clearpage

%
%

%



\clearpage
\bibliography{paper}

\end{document}